\documentclass[1p,times]{elsarticle}
\usepackage{CJKutf8}
\usepackage{multirow}
\usepackage{wrapfig}
\usepackage{booktabs}
\usepackage{rotating}
\usepackage{makecell}
\usepackage{caption} 
\usepackage{algorithm}
\usepackage{enumitem}
\usepackage{algpseudocode}
\usepackage{amsmath,amsthm,mathtools}
\usepackage{amssymb}
\usepackage{bbding}
\biboptions{sort&compress}
\usepackage{threeparttable}
\usepackage[dvipsnames]{xcolor}
\usepackage[commandnameprefix=ifneeded,defaultcolor=black]{changes}

\usepackage{mathrsfs}

\usepackage{hyperref}
\usepackage{tablefootnote}

\journal{Journal of Computer Speech and Language}

\begin{document}
\sloppy
\begin{frontmatter}

\title{Universal Speaker Embedding Free Target Speaker Extraction and Personal Voice Activity Detection}

\author[mymainaddress,mysecondaryaddress]{Bang Zeng}
\ead{bangzeng@whu.edu.cn}

\author[mymainaddress,mysecondaryaddress]{Ming Li\corref{mycorrespondingauthor}}
\cortext[mycorrespondingauthor]{Corresponding Author: Ming Li}
\ead{ming.li369@dukekunshan.edu.cn}

\address[mymainaddress]{School of Computer Science, Wuhan University, Wuhan, China}
\address[mysecondaryaddress]{Suzhou Municipal Key Laboratory of Multimodal Intelligent Systems, Digital Innovation Research Center, Duke Kunshan University, Kunshan, China}

\begin{abstract}
Determining “who spoke what and when” remains challenging in real-world applications. In typical scenarios, Speaker Diarization (SD) is employed to address the problem of “who spoke when,” while Target Speaker Extraction (TSE) or Target Speaker Automatic Speech Recognition (TSASR) techniques are utilized to resolve the issue of “who spoke what.” Although some works have achieved promising results by combining SD and TSE systems, inconsistencies remain between SD and TSE regarding both output inconsistency and scenario mismatch. To address these limitations, we propose a Universal Speaker Embedding Free Target Speaker Extraction and Personal Voice Activity Detection (USEF-TP) model that jointly performs TSE and Personal Voice Activity Detection (PVAD). USEF-TP leverages frame-level features obtained through a cross-attention mechanism as speaker-related features instead of using speaker embeddings as in traditional approaches. Additionally, a multi-task learning algorithm with a scenario-aware differentiated loss function is applied to ensure robust performance across various levels of speaker overlap. The experimental results show that our proposed USEF-TP model achieves superior performance in TSE and PVAD tasks on the LibriMix and SparseLibriMix datasets. The results on the CALLHOME dataset demonstrate the competitive performance of our model on real recordings.
\end{abstract}


\end{frontmatter}


\section{Introduction}

In real-world situations, people can easily recognize when the speaker of interest is talking and accurately understand what they are saying. Researchers have categorized this auditory ability into two tasks: Speaker Diarization (SD)~\cite{fujita2019end, Horiguchi_2020, medennikov2020target, 10094752, chen2024attention} and Blind Speech Separation (BSS)~\cite{luo2019conv, zeghidour2021wavesplit, luo2020dual, subakan2021attention, wang2023tf, zhao2024mossformer2}. SD involves determining the speech activities of multiple speakers from the input audio, answering the question of “who spoke when”. In contrast, BSS focuses on separating each speaker’s voice from a mixture of audio signals. As crucial front-end processes in speech processing, SD and BSS are widely applied in various real-world speech applications, including speaker verification~\cite{cai2024leveraging, Wang2023CAMAF, qin2024investigating} and speech recognition~\cite{yao2021wenet, prabhavalkar2023end, chang2024exploring}.

Many approaches for SD and BSS have been proposed, contributing significantly to advancements in both areas. However, these methods are generally applied to each task separately, failing to address the challenge of determining “who spoke what and when.” Due to the similarities between SD and BSS tasks, some researchers have recently proposed joint SD and BSS methods~\cite{maiti2023eend, boeddeker2024ts}. Additionally, due to the need for prior knowledge of the number of speakers in speech separation methods for practical use, some studies have focused on constructing joint models for SD and Target Speaker Extraction (TSE)~\cite{Ao_2024}. In our previous work, we have shown that cascading an SD model in front of a TSE model improves the performance of TSE~\cite{zeng2024simultaneous}.
\begin{figure*}[t!]
  \centering
  \includegraphics[width=1.0\linewidth]{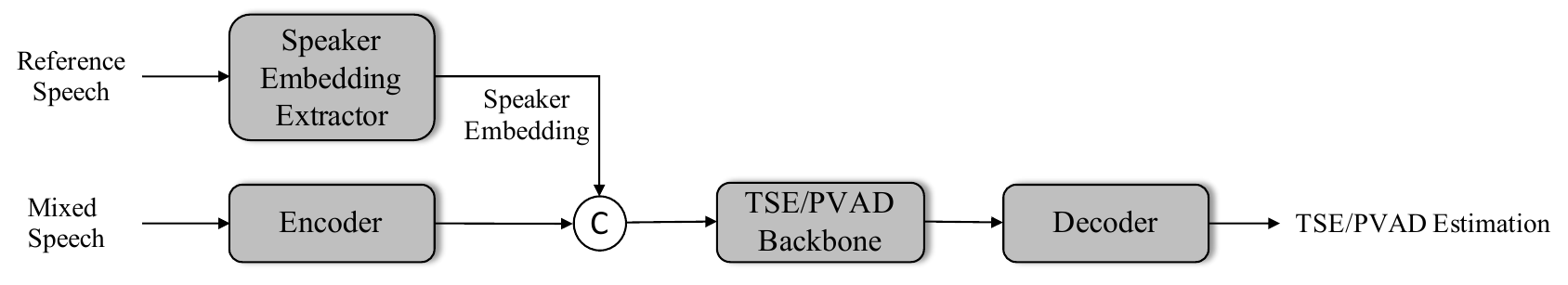}
  \caption{The diagram of a typical target speaker extraction or personal voice activity detection methods. The speaker embedding extractor is typically a pre-trained speaker recognition model. 'C' denotes the concatenation.}
  \label{fig:tytse}
\end{figure*}
However, there are some mismatches between SD and TSE, which are categorized as ‘output inconsistency’ and ‘scenario mismatch’ in~\cite{Ao_2024}. ‘Output inconsistency’ refers to the fact that an SD model outputs the speech activity boundaries for all speakers in the mixed audio. In contrast, a TSE model outputs the clean speech estimation of a specific speaker. ‘Scenario mismatch’ refers to the fact that SD is mainly applied in scenarios with less speaker overlap, whereas TSE focuses more on scenarios with a high degree of speaker overlap. 
\added{Universal Speaker Extraction and Diarization (USED)~\cite{Ao_2024} addresses output inconsistency and scenario mismatch through the embedding assignment module, supporting speech mixtures with a variable number of speakers and arbitrary overlap ratios. It leverages speaker diarization results to optimize speaker extraction while enhancing speaker diarization accuracy through the extracted speech. However, the USED approach may not be well-suited for scenarios where the number of speakers is smaller than the predefined maximum.}

In practical applications, many scenarios only require the speech activity boundaries of a specific speaker rather than those of all speakers in the mixed audio. Personal voice activity detection (PVAD)~\cite{Ding2020, Ding2022PersonalV2, zeng2024efficient} is a technique to determine the target speaker’s activity speech segments in multi-speaker scenarios. PVAD may be better suited than SD techniques for constructing joint models with TSE methods. One reason is that the outputs of PVAD and TSE models are the speech activity boundaries and clean speech predictions for a specific speaker, respectively, thus avoiding the issue of ‘output inconsistency.’ Additionally, PVAD and TSE methods do not require prior knowledge of the number of speakers in the mixed audio, ensuring consistency in the prerequisites. At last, PVAD results enhance the TSE model’s ability to suppress interfering speakers when the target speaker is silent.

As shown in Figure~\ref{fig:tytse}, typical PVAD and TSE methods first extract the target speaker’s embedding from existing reference speech using a pre-trained speaker recognition~\cite{wan2018generalized, qin2022robust, qin2024investigating} model. Subsequently, with the assistance of speaker embedding, the PVAD and TSE systems can model the target speaker’s components within the mixed audio, producing the final result. However, careful consideration is required when selecting a suitable speaker recognition model for the TSE and PVAD tasks. Moreover, these methods may not fully exploit the information contained in the reference speech. Consequently, relying solely on speaker embeddings for the TSE and PVAD tasks may not be optimal.

The speaker embedding-free framework is a viable new solution for the above problem. Our previous work, USEF-TSE~\cite{zeng2024usef}, demonstrates the effectiveness of a speaker embedding-free framework for the TSE task, achieving state-of-the-art (SOTA) performance on the WSJ0-2mix~\cite{hershey2016deep} dataset, a widely used benchmark for monaural speech separation and TSE. In another previous work~\cite{zeng2024efficient}, we show the effectiveness of a speaker embedding-free framework in the PVAD task, achieving remarkably high recall performance.

However, USEF-TSE~\cite{zeng2024usef} has only demonstrated its effectiveness in high-overlapping speech scenarios, and its performance in low-overlapping situations has yet to be validated. In~\cite{zeng2024efficient}, the model’s performance was only evaluated in short-duration speech scenarios, and its effectiveness in long-duration speech scenarios with varying overlapping ratios was not tested. Moreover, both studies were conducted independently, focusing on the TSE and PVAD tasks in isolation.

\begin{figure*}[t!]
  \centering
  \includegraphics[width=1.0\linewidth]{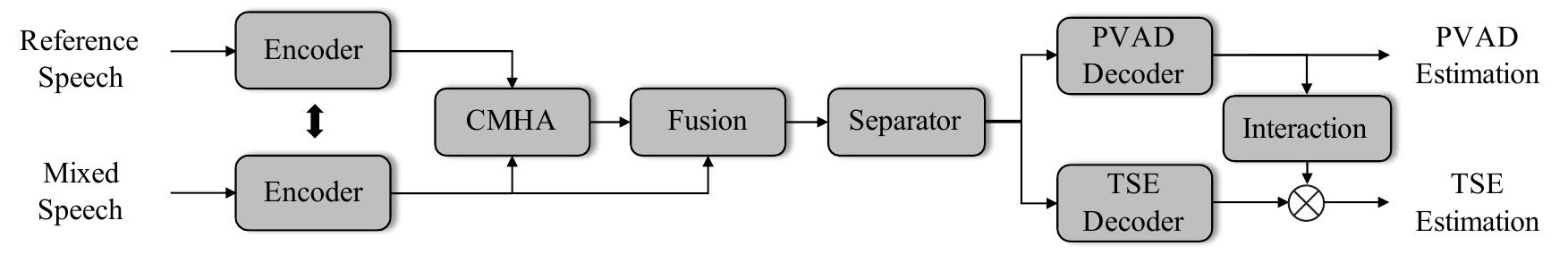}
  \caption{The diagram of the USEF-TP model.  'CMHA' denotes the cross multi-head attention. \(\otimes\) is an operation for element-wise product.}
  \label{fig:utp-sys}
\end{figure*}

To address the aforementioned issues, this paper proposes a Universal Speaker Embedding Free Target Speaker Extraction and Personal Voice Activity Detection model, referred to as USEF-TP. USEF-TP uses the same structure as USEF-TSE~\cite{zeng2024usef}. The system framework of USEF-TP is shown in Figure~\ref{fig:utp-sys}. In USEF-TP, a unified encoder processes both the reference and mixed speech signals. The encoded mixed speech features query the reference speech encoding via a cross multi-head attention module. This module outputs frame-level features that preserve the same temporal length as the encoded mixed speech features. Subsequently, USEF-TP utilizes the frame-level output as the target speaker’s characteristics and integrates it with the acoustic features of the multi-speaker input speech. After being processed by the backbone network, the fused features are passed through two deconvolution layers to generate the estimated complex spectrogram of the target speaker’s speech and the PVAD prediction values. Finally, we refine the estimated complex spectrogram using the PVAD results through an interaction module, thereby obtaining the final predicted TSE results. Although the interaction module and scene-aware loss have been validated in USED~\cite{Ao_2024}, they are specifically designed based on the joint PVAD and TSE model, aiming further to enhance the joint modeling between PVAD and TSE. This approach differs from the SD and TSE joint modeling in USED~\cite{Ao_2024}. More importantly, in contrast to USED~\cite{Ao_2024}, USEF-TP adopts a speaker-embedding-free framework, eliminating the model’s reliance on speaker embeddings and improving its performance on the TSE and PVAD tasks.

This paper is an extension of our previous work~\cite{zeng2024efficient}, and the main new contributions of this work can be summarized as follows:
\begin{itemize}
\item{ We introduce a Universal Speaker Embedding Free Target Speaker Extraction and Personal Voice Activity Detection (USEF-TP) model. The USEF-TP model performs target speaker extraction (TSE) and joint personal voice activity detection (PVAD) with the speaker-embedding-free framework}. USEF-TP addresses a unique research problem, i.e., ‘the target speaker spoke what and when.’

\item{Inspired by~\cite{Ao_2024}, we designed an interaction module to refine the estimated complex spectrogram based on the PVAD results.}

\item{We propose a multi-task learning algorithm with scenario-aware differentiated loss to optimize USEF-TP. The multi-task learning algorithm ensures temporal alignment and consistency between the PVAD and TSE results. The scenario-aware loss function can smooth the model’s performance across different levels of overlap, ensuring consistent effectiveness in both low and high-overlap scenarios.}

\item{ Our proposed USEF-TP surpasses the performance of competitive baseline systems for both TSE and PVAD on the LibriMix~\cite{cosentino2020librimix} and SparseLibriMix~\cite{cosentino2020librimix}. The results show that the USEF-TP model achieves superior performance in both highly and sparsely overlapped speech scenarios for the TSE and PVAD tasks. Moreover, the results on the CALLHOME dataset also demonstrate the effectiveness of the USEF-TP model on real recordings.}
\end{itemize}

The rest of the paper is organized as follows. Section~\ref{sec:RW} discusses the related work. The problem formulation is introduced in Section~\ref{sec:PF}. Section~\ref{sec:Methods} describes the proposed method. We introduce the experimental setup in Section~\ref{sec:ES}. Section~\ref{sec:RD} presents the experimental results and analysis. Section~\ref{sec:Conclusion} concludes the study.

\section{Related Work}
\label{sec:RW}
\subsection{Speaker Diarization and Personal Voice Activity Detection}
Early approaches to Speaker Diarization (SD)~\cite{shum2013unsupervised, sell2014speaker, park2022review} involve modular systems that use voice activity detection, speech segmentation, and clustering to determine speaker turns. These methods rely on hand-crafted features and traditional machine learning algorithms. However, they often struggle with overlapping speech and require careful tuning for each pipeline stage. The introduction of deep learning has revolutionized SD methods. Recurrent neural networks (RNNs), particularly long short-term memory (LSTM) networks, have been used to model the sequential nature of speech, leading to improved performance in detecting speaker changes. End-to-end Neural Diarization (EEND)~\cite{fujita2019end, Horiguchi_2020, 9003959} systems and Target Speaker Voice Activity Detection (TS-VAD)~\cite{10094752, 10095185} further simplified the process by treating SD as a multi-label classification problem. EEND has become an effective method for consolidating the multiple stages of SD into a single, unified model. These systems are designed to predict speaker activity segments from raw audio directly. By jointly optimizing all stages of the SD process, EEND systems offer improved performance over traditional pipeline-based approaches. TS-VAD is a specialized VAD technique aimed at identifying the speech activity of a group of speakers within a multi-speaker environment. This approach is particularly valuable in applications like meeting transcription, where the number of speakers are high and the duration is long. Recent methods have leveraged sequence-to-sequence prediction models to enhance the efficiency of TS-VAD, particularly in challenging acoustic environments~\cite{cheng2024multi}.

Unlike SD, Personal Voice Activity Detection (PVAD)~\cite{Ding2020, Ding2022PersonalV2, zeng2024efficient} models enable the detection of speech from a specific speaker while ignoring other speakers or background noise within a multi-speaker environment. This method benefits personal assistants and smart home devices, where the target speaker is known as a priori. PVAD has been approached by using speaker embeddings, such as I-vectors~\cite{he2021target} or X-vectors~\cite{snyder2018x}, to detect the voice activity of a known speaker in a recording. However, the speaker embedding extraction process can be sensitive to noise and other environmental factors, leading to false positives in detecting the target speaker. Recent work has introduced the speaker embedding free PVAD framework, eliminating the need for an additional speaker recognition model to extract speaker embeddings for a specific individual~\cite{zeng2024efficient}.  

Despite the progress, challenges remain in handling highly overlapping speech for joint SD and PVAD. It remains an area where further improvement is needed, especially in real applications.

\subsection{Speech Separation and Target Speaker Extraction}
Blind Speech Separation (BSS) involves separating the voice of all sources. Traditional BSS approaches, such as Non-negative Matrix Factorization (NMF)~\cite{schmidt2006single, cichocki2006new} and Computational Auditory Scene Analysis (CASA)~\cite{lyon1983computational,wang2006computational}, typically use spectro-temporal masking to isolate each speaker’s component from mixed speech. More recent methods have leveraged deep learning to improve separation quality. Deep neural network-based speech separation algorithms, such as Deep Clustering (DC)~\cite{hershey2016deep,Isik_2016}, Deep Attractor Network (DANet)~\cite{chen2017deep,luo2018speaker}, and Permutation Invariant Training (PIT)~\cite{yu2017permutation}, have significantly improved the performance of speech separation tasks. However, traditional time-frequency domain methods for BSS face a notable limitation: the inability to accurately reconstruct the phase of clean speech. The introduction of time-domain BSS networks (TasNet)~\cite{luo2018tasnet} addressed this issue. Several time-domain methods, such as Dual-Path RNN (DPRNN)~\cite{luo2020dual}, SepFormer~\cite{subakan2021attention}, Mossformer~\cite{zhao2023mossformer, zhao2024mossformer2} further enhance the speech separation model’s performance. The time-frequency domain model TF-GridNet~\cite{wang2023tf} have achieved remarkable progress. Despite these advancements, many BSS techniques still require prior knowledge of the number of speakers in the mixture, which is not always available in real-world applications. This problem poses significant challenges for deploying these BSS solutions in practical settings.

Target Speaker Extraction (TSE)~\cite{wang19h_interspeech, vzmolikova2019speakerbeam, li20p_interspeech, Zhang2020XTaSNetRA, xu2020spex} can extract a specific speaker's voice from the mixed audio by utilizing an auxiliary reference speech of the target speaker. A typical TSE model relays on the target speaker embedding from a pre-trained~\cite{liu2023x,hao2024x} or joint-learned~\cite{ge2020spex+, ge2021multi, wang2021neural} speaker embedding extractor. \added{Because embedding extractors can be fine-tuned or adapted to new domains to improve performance, the multi-task joint training approaches adopted in ~\cite{ge2020spex+, ge2021multi, wang2021neural} have significantly enhanced the model's performance.} However, the aim of training this extractor is usually to maximize speaker recognition performance. Moreover, these methods may only partially utilize some of the information in the reference speech. Recent developments have investigated embedding-free methods that utilize frame-level acoustic features from reference speech, bypassing the need for speaker embeddings. The VE-VE framework~\cite{yang2023target} employs an RNN-based voice extractor that captures speaker characteristics using RNN states instead of speaker embeddings. This approach effectively handles the feature fusion challenge, though it is constrained to RNN-based extraction networks. SEF-Net~\cite{zeng2023sef}, CIENet~\cite{yang2024target} and USEF-TSE~\cite{zeng2024usef} employ attention mechanisms to interact with the encoder of both the reference and mixed signals. These methods can guide more effective extraction by leveraging contextual information directly.

While these approaches have shown promising results, they are typically optimized for scenarios with heavily overlapped speech, limiting their effectiveness in less overlapping conditions.

\subsection{Joint Speech Separation or Target Speaker Extraction with Speaker Diarization}
Given the alignment in objectives between the TSE (or BSS) and SD tasks, recent studies have combined these tasks for multi-task training. EEND-SS~\cite{maiti2023eend} introduces a framework that jointly performs SD, BSS, and speaker counting for a flexible number of speakers. It proposes a multiple 1D convolutional layer architecture for estimating separation masks and a fusion technique to refine separated speech signals using SD information. However, this fusion technique is only performed in the inference stage. USED~\cite{Ao_2024} introduces a unified approach designed to address the challenges of TSE and SD in multi-talker scenarios. The model employs an embedding assignment module to manage outputs for a flexible number of speakers and a multi-task interaction module to leverage complementary information from both tasks. The USED~\cite{Ao_2024} model represents a significant advancement in the field, offering a comprehensive solution for "who spoke what and when." USED incorporates an embedding assignment module that dynamically adjusts the number of speaker outputs based on detected speakers, reducing output inconsistencies. However, it still relies on a preset maximum number of speakers during training. If the number of speakers in an evaluation scenario exceeds this limit, the model’s ability to assign embeddings accurately and separate speakers effectively may be affected.

\section{Problem Formulation}
\label{sec:PF}
In real-world speech scenarios, a conversation may involve multiple speakers’ voices along with background noise and reverberation:
\begin{equation}
  \boldsymbol{m} = \sum_{i=1}^{M}{\boldsymbol{x}_{i} + \boldsymbol{n}}
  \label{eq1}
\end{equation}
where \(\boldsymbol{m}\) \(\in\) \(\mathbb{R}^{1 \times T_{m}}\) denotes the mixed speech, \(T_{m}\) is the length of the mixed speech. \(\boldsymbol{x}_{i}\) denotes the speech of the \(i^{th}\) speaker. \(\boldsymbol{n}\) denotes the additive noise. \(M\) is the number of speakers in the mixed speech.

The objective of the USEF-TP model is to detect the speech activity segments of the target speaker and extract the target speaker’s voice from the mixed audio:
\begin{equation}
  \boldsymbol{\hat{x}}_{tgt}, \boldsymbol{\hat{p}}_{tgt} = \mathscr{M}(\boldsymbol{m}, \boldsymbol{r}) = \mathscr{M}(\boldsymbol{x}_{tgt} + \sum_{i=1}^{M-1}{\boldsymbol{x}_{i} + \boldsymbol{n}}, \boldsymbol{r})
  \label{eq2}
\end{equation}
where \(\boldsymbol{\hat{x}}_{tgt}\) denotes the extracted speech of the target speaker. \(\boldsymbol{\hat{p}}_{tgt}\) denotes the VAD precision of the target speaker. \(\mathscr{M}(\cdot)\) denotes operations in the USEF-TP model. \(\boldsymbol{x}_{tgt}\) represents the speech component of the target speaker within \(\boldsymbol{m}\). \(\boldsymbol{r}\) \(\in\) \(\mathbb{R}^{1 \times T_{r}}\) represents the reference speech, \(T_{r}\) is the length of the reference speech.
\begin{figure*}[t!]
  \centering
  \includegraphics[width=0.6\linewidth]{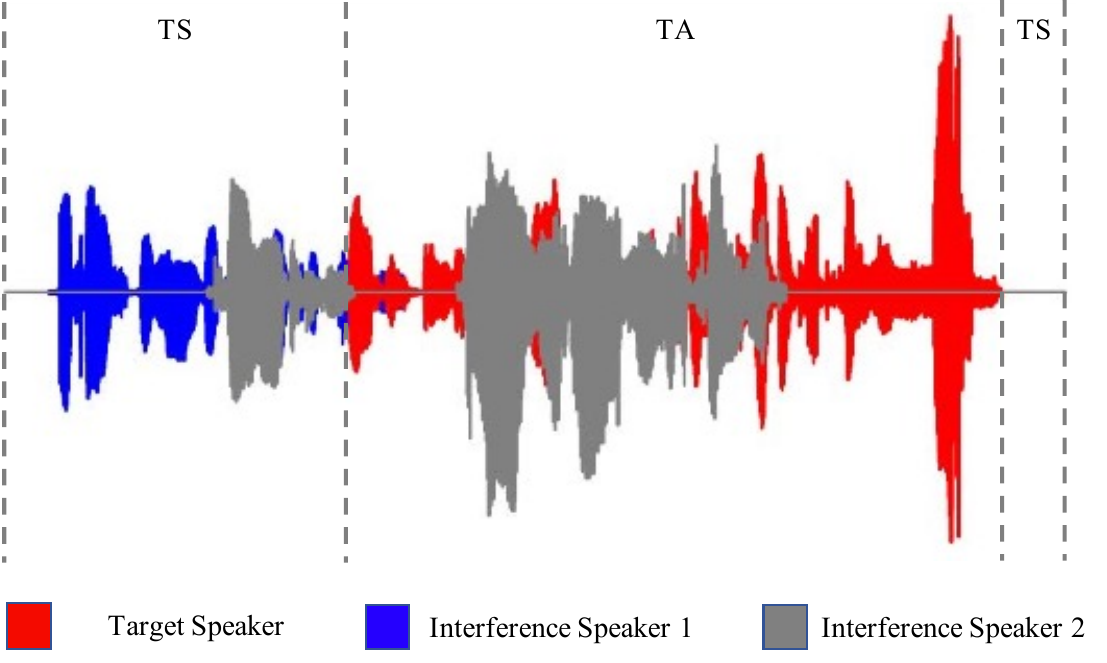}
  \caption{The diagram of different scene clips from a mixed audio recording. 'TA' denotes Target epaker Active. 'TS' denotes Target speaker Silence.}
  \label{fig:TATS}
\end{figure*}

It is worth noting that the expected output of the model \(\boldsymbol{\hat{x}}_{tgt}\) will vary depending on whether the target speaker is present or absent in the speech clip. In~\cite{pan2022usev}, speech clips are classified into four classes: QQ, QS, SS, and SQ. In the QQ scenario, neither the target nor the interfering speakers are active. The QS scenario represents a case where the target speaker is silent while the interfering speakers are active. Conversely, the SQ scenario occurs when the target speaker is speaking, and the interfering speakers remain silent. Lastly, the SS scenario involves both the target and interfering speakers being active simultaneously. Nevertheless, this study focuses on tasks related to only one target speaker. As a result, we simplify the previous four classifications into two categories: Target-speaker Active (TA) and Target-speaker Silence (TS). Figure~\ref{fig:TATS} illustrates an example of different scene clips from a mixed audio recording. In the TA scenario, the target speaker is active, encompassing both the SS and SQ scenarios. The expected TSE output of the model is to accurately restore a prediction that closely aligns with the target speaker’s speech components. Conversely, in the TS scenario, the target speaker is quiet, which includes the QS and QQ scenarios. In this scenario, the expected output of TSE is 0, indicating no speech activity from the target speaker:
\begin{equation}
  \mathscr{M}(\boldsymbol{m}, \boldsymbol{r})_{tse} = \begin{cases}
  \boldsymbol{\hat{x}}_{tgt} \to\ \boldsymbol{x}_{tgt}, SC \in TA \\
  \boldsymbol{\hat{x}}_{tgt} \to 0, SC \in TS
  \end{cases}
  \label{eq3}
\end{equation}
where ‘SC’ denotes the speech clip. \(\mathscr{M}(\cdot)_{tse}\) denotes the TSE output of the USEF-TP model.

\section{Methods}
\label{sec:Methods}

\begin{figure*}[t!]
  \centering
  \includegraphics[width=1.0\linewidth]{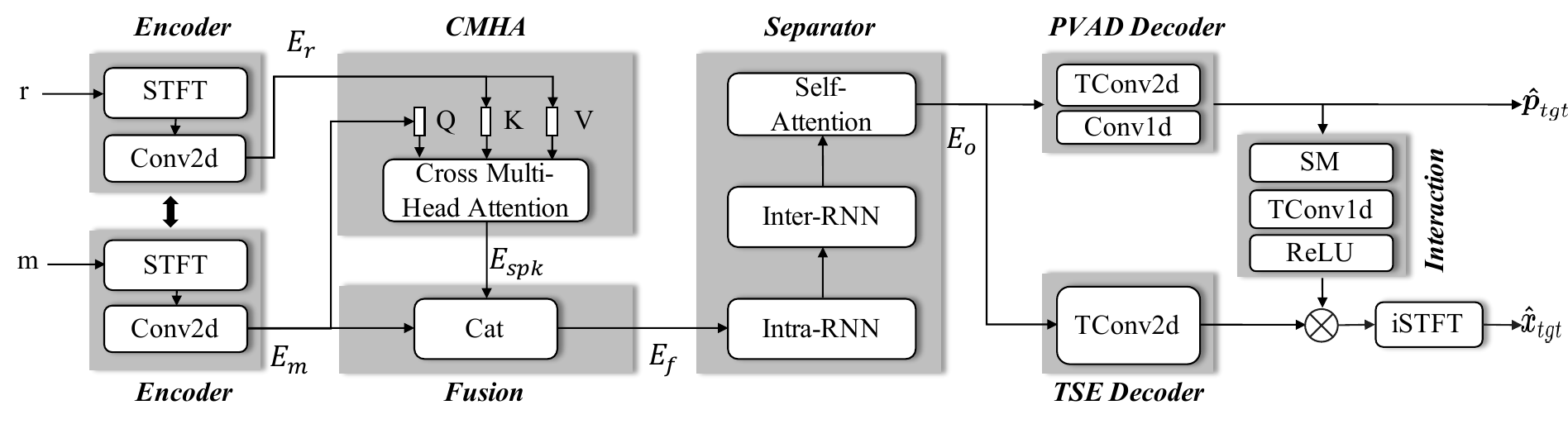}
  \caption{The diagram of USEF-TP model. 'm' and 'r' denote the mixed speech and reference speech, respectively. We use two weight sharing encoder to process the mixed and reference speech separately. \(\otimes\) is an operation for element-wise product. The Separator’s parameters are set identically to those of the TF-GridNet approach.}
  \label{fig:usef-tp}
  \vspace{-0.2cm}
\end{figure*}

The architecture of the USEF-TP model is illustrated in Figure~\ref{fig:usef-tp}. Our proposed USEF-TP model consists of seven modules: the encoder, Cross Multi-Head Attention (CMHA) module, fusion module, separator, TSE decoder, PVAD decoder, and interaction module. This section will provide a detailed description of the USEF-TP model.

\subsection{Encoder}
Motivated by TF-GridNet~\cite{wang2023tf}, USEF-TP uses two weight-sharing convolutional encoders to process the mixed speech \(\boldsymbol{m}\) \(\in\) \(\mathbb{R}^{B \times 1 \times T_{m}}\)and the reference speech \(\boldsymbol{r}\) \(\in\) \(\mathbb{R}^{B \times 1 \times T_{r}}\). \(T_{m}\) and \(T_{r}\) denote the length of the mixed and reference speech, respectively. Similar to USEF-TSE~\cite{zeng2024usef}, This encoder can be a Short-Time Fourier transform (STFT) or a one-dimensional convolution, depending on whether the model works in the time or time-frequency (T-F)  domain. In this work, the USEF-TP model operates in the T-F domain:
\begin{equation}
  \boldsymbol{m}_{RI} = STFT(\boldsymbol{m})
  \label{eq4}
\end{equation}
\begin{equation}
  \boldsymbol{r}_{RI} = STFT(\boldsymbol{r})
  \label{eq5}
\end{equation}
where \(STFT(\cdot)\) refers to STFT operation. \(\boldsymbol{m}_{RI}\) \(\in\) \(\mathbb{R}^{B \times 2 \times F \times L_{m}}\) and \(\boldsymbol{r}_{RI}\) \(\in\) \(\mathbb{R}^{B \times 2 \times F \times L_{r}}\) represent the stacked real and imaginary components of the STFT features for the mixed and reference speech, respectively. \(B\) denotes the batch size, \(F\) is the feature dimension for each T-F unit. \(L_{m}\) and \(L_{r}\) is the number of frames of \(\boldsymbol{m}_{RI}\) and \(\boldsymbol{r}_{RI}\), respectively.
\begin{equation}
  \boldsymbol{E}_{m} = Conv2d(\boldsymbol{m}_{RI})
  \label{eq6}
\end{equation}
\begin{equation}
  \boldsymbol{E}_{r} = Conv2d(\boldsymbol{r}_{RI})
  \label{eq7}
\end{equation}
where \(Conv2d(\cdot)\) refers to 2D convolutions. \(\boldsymbol{E}_{m}\) \(\in\) \(\mathbb{R}^{B \times C \times F \times L_{m}}\) and \(\boldsymbol{E}_{r}\) \(\in\) \(\mathbb{R}^{B \times C \times F \times L_{r}}\) denotes the encoder results for the mixed and reference speech, respectively. The purpose of \(Conv2d(\cdot)\) is to transform the number of feature channels from the original 2 to \(C\). 
\subsection{CMHA module}
The encoder outputs of \(\boldsymbol{m}\) and \(\boldsymbol{r}\) are passed into the CMHA module, which employs a cross multi-head attention mechanism:
\begin{equation}
  \boldsymbol{E}_{spk} = {CMHA}(q=\boldsymbol{E}_{m}; k,v=\boldsymbol{E}_{r})
  \label{eq8}
\end{equation}
where \(\boldsymbol{E}_{m}\) \(\in\) \(\mathbb{R}^{B \times C \times F \times L_{m}}\) and \(\boldsymbol{E}_{r}\) \(\in\) \(\mathbb{R}^{B \times C \times F \times L_{r}}\) represent the encoder outputs of the mixed speech and reference speech, respectively. \(\boldsymbol{E}_{spk} \in \mathbb{R}^{B \times C \times F \times L_{m}}\) represents the output of the CMHA module. CMHA denotes the Cross Multi-Head Attention operation. To align with the separator, the USEF-TP model uses the attention module from TF-GridNet~\cite{wang2023tf} as its CMHA module.

The CMHA module of the USEF-TP model uses the mixed speech encoding \(\boldsymbol{E}_{m}\) as the query and the reference speech encoding \(\boldsymbol{E}_{r}\) as the key and value. Through the Equation (\ref{eq8}), the CMHA module produces a frame-level feature \(\boldsymbol{E}_{spk}\), which with the same length as \(\boldsymbol{E}_{m}\). In this way, the USEF-TP model allows the lengths of the mixed speech input \(L_{m}\) and the reference speech input \(L_r\) to be different.

Unlike the traditional approach shown in Figure~\ref{fig:tytse}, The USEF-TP model does not rely on an additional pre-trained speaker recognition model to extract the target speaker’s embedding. Instead, it uses the output of the CMHA module as the target speaker's attributes. Compared to fixed-dimensional speaker embeddings, this frame-level feature \(\boldsymbol{E}_{spk}\) not only retains target speaker information but also incorporates context-dependent information, which facilitates improved TSE or PVAD performance~\cite{zeng2024usef}.

\subsection{Fusion Module}
After the CMHA module, the USEF-TP model uses a fusion module to combine \(\boldsymbol{E}_{spk}\) and the mixed speech encoding \(\boldsymbol{E}_{m}\):
\begin{equation}
  \boldsymbol{E}_{f} = {F}(\boldsymbol{E}_{m}, \boldsymbol{E}_{spk})
  \label{eq9}
\end{equation}
where \(\boldsymbol{E}_{f}\) denotes the output of the fusion module. \(F(\cdot)\) denotes the feature fusion operation. 

There are generally two approaches to feature fusion. One is to directly concatenate features along the channel dimension, resulting in a fused feature \(\boldsymbol{E}_{f}\) \(\in\) \(\mathbb{R}^{B \times 2C \times F \times L_{m}}\). The other approach is to use Feature-wise Linear Modulation (FiLM)~\cite{perez2018film}:
\begin{equation}
  \boldsymbol{E}_{f} = FiLM(\boldsymbol{E}_{m}, \boldsymbol{E}_{spk})
  \label{eq10}
\end{equation}
\begin{equation}
  FiLM(\boldsymbol{E}_{m}, \boldsymbol{E}_{spk}) = \gamma(\boldsymbol{E}_{spk}) \cdot \boldsymbol{E}_{m} + \beta(\boldsymbol{E}_{spk})
  \label{eq11}
\end{equation}
Here, \(\boldsymbol{E}_{f} \in \mathbb{R}^{B \times C \times F \times L_{m}}\) denotes the fused features. The scaling and shifting vectors in FiLM are denoted by  \(\gamma(\cdot)\) and \(\beta(\cdot)\), respectively.

In the USEF-TP model, we use concatenation as the feature fusion method. The fused features \(\boldsymbol{E}_{f}\) \(\in\) \(\mathbb{R}^{B \times 2C \times F \times L_{m}}\), are then passed into the separator.

\subsection{Separator}
The USEF-TP model uses TF-GridNet as the backbone of the separator. The separator comprises three modules: the full-band, the sub-band, and the cross-frame self-attention module. The separator first applies zero-padding and unfold operation to the feature \(\boldsymbol{E}_{f}\):
\begin{equation}
 \boldsymbol{E}^\prime_{f} = P(\boldsymbol{E}_{f})
  \label{eq12}
\end{equation}
\begin{equation}
 \boldsymbol{U}_{full} = U(RS(\boldsymbol{E}^\prime_{f}))
  \label{eq13}
\end{equation}
where \(\boldsymbol{E}^\prime_{f}\) \(\in\) \(\mathbb{R}^{B \times 2C \times F^\prime \times L_{m}^\prime}\) represents the zero-padded version of \(\boldsymbol{E}_{f}\). \(P(\cdot)\) denotes the zero-padding operation. The reshape operation, denoted as \(RS(\cdot)\), reshapes \(\boldsymbol{E}^\prime_{f}\) into the shape \(\mathbb{R}^{(B \times L_{m}^\prime) \times F^\prime \times 2C}\). Here, \(F^\prime = \lceil\frac{F-ks}{hs}\rceil \times hs + ks\). The function \(U(\cdot)\) corresponds to the torch.unfold function, with \(\boldsymbol{U}_{full}\) \(\in\) \(\mathbb{R}^{(B \times L_{m}^\prime) \times (\frac{F^\prime-ks}{hs}+1) \times (2C \times ks)}\) as its output. The parameters \(ks\), \(hs\) are the kernel size and stride of the torch.unfold function, respectively.

After the unfolding process, the features pass through the intra-frame full-band module and the sub-band module sequentially, similar to TF-GridNet~\cite{wang2023tf}, generating \(\boldsymbol{E}_{full}\) and  \(\boldsymbol{E}_{sub}\) \(\in\) \(\mathbb{R}^{B \times 2C \times F^\prime \times L_{m}^\prime}\). The output of the sub-band module \(\boldsymbol{E}_{sub}\) is subsequently fed into the attention module. Before this, the zero-padding in the sub-band results must be removed to restore the length to that of the original input:
\begin{equation}
  \boldsymbol{E}_{cross} = Cross(\boldsymbol{E}_{sub}[:,:,F,L_{m}])
  \label{eq17}
\end{equation}
\begin{equation}
  \boldsymbol{E}_{o} = \boldsymbol{E}_{cross} + \boldsymbol{E}_{sub}[:,:,F,L_{m}]
  \label{eq18}
\end{equation}
where \(Cross(\cdot)\) refers to the operations in the cross-frame self-attention module, which applies the multi-head attention on the input. In this way, each T-F unit can attend to other relevant units.  \(\boldsymbol{E}_{Cross}\), \(\boldsymbol{E}_{o}\) \(\in\) \(\mathbb{R}^{B \times 2C \times F \times L_{m}}\) are the outputs of the cross-frame self-attention module and the separator, respectively.

\subsection{TSE and PVAD Decoder}
\subsubsection{TSE Decoder}
In the TSE Decoder of the USEF-TP model, a 2D transposed convolution layer is applied to adjust the number of channels in \(\boldsymbol{E}_{o}\) back to 2:
\begin{equation}
  \boldsymbol{D}_{tse} = TConv2d_{tse}(\boldsymbol{E}_{o})
  \label{eq19}
\end{equation}
where \(\boldsymbol{D}_{tse}\) \(\in\)  \(\mathbb{R}^{B \times 2 \times F \times L_{m}}\) denotes the output of the TSE Decoder. \(TConv2d_{tse}(\cdot)\) denotes the 2D transposed convolution layer in the TSE Decoder.

\subsubsection{PVAD Decoder}
In the PVAD Decoder of the USEF-TP model,  a 2D transposed convolution layer is applied to adjust the number of channels in \(\boldsymbol{E}_{o}\) back to 1:
\begin{equation}
  \boldsymbol{D}_{pvad} = TConv2d_{pvad}(\boldsymbol{E}_{o})
  \label{eq20}
\end{equation}
where \(\boldsymbol{D}_{pvad}\) \(\in\)  \(\mathbb{R}^{B \times 1 \times F \times L_{m}}\) denotes the output of the 2D transposed convolution layer. \(TConv2d_{pvad}(\cdot)\) denotes the 2D transposed convolution layer in the PVAD decoder. Then,  the PVAD decoder applies a 1D convolution layer transforms the feature dimension of \(\boldsymbol{D}_{pvad}\) to 1, while also adjusting its length:
\begin{equation}
  \boldsymbol{\hat{p}}_{tgt} = Conv1d_{pvad}(\boldsymbol{D}_{pvad})
  \label{eq21}
\end{equation}
where \(\boldsymbol{\hat{p}}_{tgt}\) \(\in\)  \(\mathbb{R}^{B \times 1 \times L_{pvad}}\) denotes the estimation of the PVAD. \(L_{pvad}\) denotes the length of the PVAD labels. \(Conv1d_{pvad}(\cdot)\) denotes the 1D convolution layer in the PVAD decoder.

\subsection{Interaction Module}
Since the training objectives of the TSE and PVAD tasks are highly aligned, the results of these tasks can enhance each other. We can leverage this consistency to refine the results of TSE. Inspired by the USED~\cite{Ao_2024} model, we added an Interaction module to the USEF-TP model. The goal of the Interaction module is to feed the results of PVAD back into the TSE predictions:
\begin{equation}
  \boldsymbol{\hat{p}^\prime}_{tgt} = ReLU(TConv1d_{IM}(SM(\boldsymbol{\hat{p}}_{tgt})))
  \label{eq22}
\end{equation}
\begin{equation}
  \boldsymbol{\hat{x}}_{tgt} = iSTFT(\boldsymbol{D}_{tse} \cdot [\boldsymbol{\hat{p}^\prime}_{tgt},\boldsymbol{\hat{p}^\prime}_{tgt}])
  \label{eq23}
\end{equation}
where \(\boldsymbol{\hat{p}^\prime}_{tgt}\) \(\in\) \(\mathbb{R}^{B \times 1 \times L_{m}}\) denotes the output of the interaction module. \(ReLu(\cdot)\) and \(SM(\cdot)\) denote the Rectified Linear Unit (ReLU) and softmax function, respectively. \(TConv1d_{IM}(\cdot)\) denotes the 1D transposed convolution layer in the interaction module. The purpose of the 1D transposed convolution is to up-sample \(\boldsymbol{\hat{p}}_{tgt}\), so that the length of \(\boldsymbol{\hat{p}^\prime}_{tgt}\) matches the length of \(\boldsymbol{D}_{tse}\). Next, we repeat \(\boldsymbol{\hat{p}^\prime}_{tgt}\) along the channel dimension and get \([\boldsymbol{\hat{p}^\prime}_{tgt},\boldsymbol{\hat{p}^\prime}_{tgt}]\) \(\in\) \(\mathbb{R}^{B \times 2 \times 1 \times L_{m}}\). Then, the element-wise multiplication is performed between \(\boldsymbol{D}_{tse}\) \(\in\)  \(\mathbb{R}^{B \times 2 \times F \times L_{m}}\) and \([\boldsymbol{\hat{p}^\prime}_{tgt},\boldsymbol{\hat{p}^\prime}_{tgt}]\). Finally, the inverse STFT (iSTFT) is applied to obtain the final TSE prediction \(\boldsymbol{\hat{x}}_{tgt}\) \(\in\) \(\mathbb{R}^{B \times 1 \times T_{m}}\).

\subsection{Loss Function}
The loss function consists of TSE loss and PVAD loss:
\begin{equation}
  \mathscr{L} = \lambda_{1} \cdot \mathscr{L}_{TSE} + \lambda_{2} \cdot \mathscr{L}_{PVAD}
  \label{eq24}
\end{equation}
where \(\mathscr{L}_{TSE}\) and \(\mathscr{L}_{PVAD}\) denote the TSE and PVAD loss, respectively. \(\lambda_{1}\) and \(\lambda_{2}\) represent the weights for TSE and PVAD loss, respectively.

For the TSE loss, we use a scene-aware loss function~\cite{Ao_2024}. In Section~\ref{sec:PF}, we categorized the speech segments in the mixed audio into TA and TS scenarios. In TA, the target speaker is active, while in TS, the target speaker is silent. For TA clips, we use the scale-invariant signal-to-distortion ratio (SI-SDR)~\cite{le2019sdr} as the objective for the TSE estimation:
\begin{equation}
\begin{cases}
 \boldsymbol{s}_{T} = \frac{<\hat{\boldsymbol{s}},\boldsymbol{s}>\boldsymbol{s}}{||\boldsymbol{s}||^2}\\ 
\boldsymbol{s}_{E} = \hat{\boldsymbol{s}} - \boldsymbol{s}_{T} \\
\mathscr{L}_{SI-SDR} = -10\lg{\frac{||\boldsymbol{s}_{T}||^2}{||\boldsymbol{s}_{E}||^2}}
\end{cases}
\label{eq25}
\end{equation}
where \(\hat{\boldsymbol{s}}\) \(\in\) \(\mathbb{R}^{1 \times T}\) represents the estimated target speaker speech, while \(\boldsymbol{s}\) \(\in\) \(\mathbb{R}^{1 \times T}\) represents the clean target speech. \(<\boldsymbol{s},\boldsymbol{s}>\) denotes the power of the signal \(\boldsymbol{s}\). For TS clips, we use the power loss as the objective for TSE estimation:
\begin{equation}
\mathscr{L}_{p} = 10\lg{(|||\boldsymbol{s}||^2 - ||\boldsymbol{\hat{s}}||^2| + \epsilon)}
\label{eq26}
\end{equation}
%
Unlike USEV~\cite{pan2022usev}, we do not directly use power but instead adopt an energy-based loss ( \(\mathscr{L}_{p} = 10\lg{(||\boldsymbol{\hat{s}}||^2 + \epsilon)}\) ) as the loss function. This change is mainly motivated by two considerations: first, during training, the energy in the TS segments of the target speaker’s clean speech labels is nearly zero, so Equation (\ref{eq26}) can effectively ensure that the energy in the TS segments of the TSE estimation approaches zero. Second, when the energy of the clean target speech in the labeled TS segments is non-zero (due to label noise), using energy loss as the loss function can help preserve the overall SI-SDR of the TSE estimation as much as possible. The scene-aware TSE loss \(\mathscr{L}_{TSE}\) can be redefine as follows:
\begin{equation}
\mathscr{L}_{TSE} = \alpha_{1} \cdot \mathscr{L}_{SI-SDR} + \alpha_{2} \cdot \mathscr{L}_{p}
\label{eq27}
\end{equation}
where \(\alpha_{1}\) and \(\alpha_{2}\) are the weights for the \(\mathscr{L}_{SI-SDR}\) and \(\mathscr{L}_{p}\), respectively. We use the binary cross-entropy loss to optimize the model for the PVAD task:
\begin{equation}
\mathscr{L}_{PVAD} = BCE(\boldsymbol{\hat{p}}_{tgt}, \boldsymbol{p}_{tgt})
\label{eq28}
\end{equation}
where \(BCE(\cdot)\) denotes the binary cross-entropy loss function. \(\boldsymbol{p}_{tgt}\) and \(\boldsymbol{\hat{p}}_{tgt}\) represent the ground-truth VAD label of the target speaker and PVAD predicted, respectively.

\section{Experimental Setup}
\label{sec:ES}
\subsection{Datasets}
\subsubsection{LibriMix}
The LibriMix~\cite{cosentino2020librimix} dataset is derived from LibriSpeech~\cite{7178964} and WHAM!'s noises~\cite{Wichern2019WHAM}. LibriMix is widely used for speech separation, TSE, and speaker diarization tasks. It primarily includes two subsets: Libri2Mix and Libri3Mix. The dataset offers four variations based on two sampling rates (16 and 8 kHz) and two modes (min and max). In min mode, the mixture ends with the shortest utterance, while in max mode, the shortest utterance is padded to match the length of the longest one. We follow ~\cite{Ao_2024} in our experiments to combine the Libri2mix 100h and Libri3mix 100h datasets as the training set. Only the “max” version at a 16 kHz sampling rate is used in the training stage. We evaluate our proposed model on the max modes.

Same as in ~\cite{Ao_2024}, we follow the scripts~\footnotemark[1]\footnotetext[1]{\url{https://github.com/msinanyildirim/USED-splits}} to slightly adjust the data split for training and validation and the scripts~\footnotemark[2]\footnotetext[2]{\url{https://github.com/gemengtju/L-SpEx/tree/main/data}} to prepare the reference speech of the target speaker.

\subsubsection{SparseLibriMix}
In addition to evaluating the model on the test set of LibriMix, we also tested the model’s performance on the SparseLibriMix~\cite{cosentino2020librimix} datasets. Compared to LibriMix, the SparseLibriMix dataset moves towards more realistic, conversation-like scenarios. SparseLibriMix is a sparsely overlapping version of LibriMix. This dataset encompasses more lifelike mixture scenarios, with overlap ratios varying from 0 to 1.0. 

\subsubsection{CALLHOME}
We test the effectiveness of the USEF-TP model on the CALLHOME dataset to evaluate the SD performance of our proposed model on real recordings. We use libri2mix and libri3mix, generated from the LibriSpeech train-clean-100 subset, as the pre-training dataset, totaling approximately 98 hours of data. We then fine-tuned the pre-trained model on the CALLHOME Part 1 dataset and evaluated it on CALLHOME Part 2. It is important to note that we directly use a raw speech segment for enrollment, meaning that higher audio quality benefits our model. In this work, We use the SD results of Diaper~\footnotemark[3]\footnotetext[3]{\url{https://github.com/BUTSpeechFIT/DiaPer}}~\cite{landini2024diaper} to extract the reference speech.

\subsection{Network Configuration}
\subsubsection{PVAD 1.0 and PVAD 2.0}
We use the PVAD 1.0~\cite{Ding2020} and PVAD 2.0~\cite{Ding2022PersonalV2} as the baseline models for the PVAD task. The PVAD 1.0 is a typical PVAD model with concatenation as the speaker embedding fusion module. To align the model complexity with that of the USEF-TP model, we use a 4-layer LSTM network with 256 neurons as the backbone for the PVAD 1.0 model. PVAD 2.0 is a conformer-based model with FiLM as the speaker embedding fusion module. Similarly, to ensure that the PVAD 2.0 model has parameters consistent with the USEF-TP model, we adjusted the size of PVAD 2.0 accordingly. The conformer-based PVAD 2.0 models have 4 Conformer~\cite{gulati20_interspeech} layers in the block, each having a dimension of 256, attention head of 8, feed-forward dimension of 1024, causal 7 × 7 convolution kernel, and 31 left-context. We set the number of the conformer block to 2.
\begin{table}[]
\centering
\caption{The parameter configuration of the USEF-TP model.}
  \label{tab:hyp}
\begin{tabular}{c|c}
\toprule
\textbf{Description}                                                                                                                                          & \textbf{Configuration} \\ \hline
\begin{tabular}[c]{@{}c@{}}
The number of the attention layers, attention heads,  \\ and dimension of the feed-forward network in the CMHA module\end{tabular} & 1, 4, 512          \\ \hline
\begin{tabular}[c]{@{}c@{}}Kernel size, stride size, input , \\ and output dimensions for the Conv2d in the Encoder\end{tabular}                              & (3,3), 1, 2, 128   \\ \hline
Kernel and stride size for Unfold                                                                                                                             & 1, 1               \\ \hline
Number of the TF-GridNet blocks                                                                                                                               & 6                  \\ \hline
\begin{tabular}[c]{@{}c@{}}Number of hidden units of BLSTMs \\ in the TF-GridNet blocks\end{tabular}                                                          & 256                \\ \hline
\begin{tabular}[c]{@{}c@{}}Kernel size, stride size, input , \\ and output dimensions for the TConv2d in the TSE Decoder\end{tabular}                         & (3,3), 1, 256, 2   \\ \hline
\begin{tabular}[c]{@{}c@{}}Kernel size, stride size, input , \\ and output dimensions for the TConv2d in the PVAD Decoder\end{tabular}                        & (3,3), 2, 256, 1   \\ \hline
\begin{tabular}[c]{@{}c@{}}Kernel size, stride size, input , \\ and output dimensions for the Conv1d in the PVAD Decoder\end{tabular}                         & 2, 1, 161, 1       \\ \hline
\begin{tabular}[c]{@{}c@{}}Kernel size, stride size, input , \\ and output dimensions for the TConv1d in the Interaction module\end{tabular}                  & 2, 1, 1, 1        \\ 
\bottomrule
\end{tabular}
\end{table}
\subsubsection{USEF-TP}
\begin{figure*}[t!]
  \centering
  \includegraphics[width=1.0\linewidth]{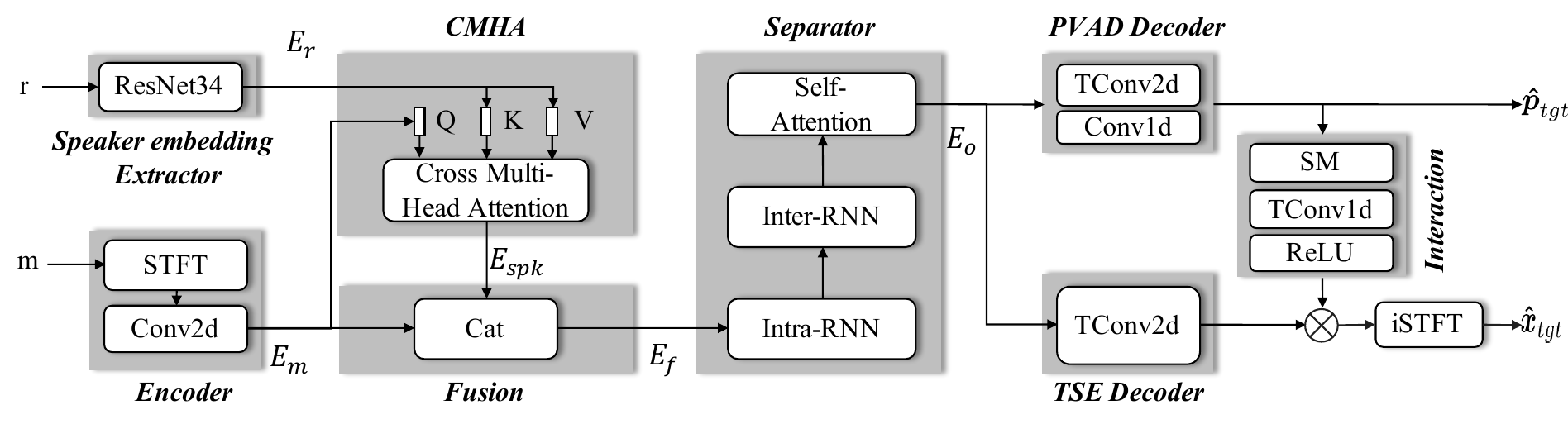}
  \caption{The diagram of SEB-TP model. 'm' and 'r' denote the mixed speech and refernece speech, respectively. \(\boldsymbol{E}_{r}\) denotes the target speaker embedding. \(\otimes\) is an operation for element-wise product. The Separator’s parameters are set identically to those of the TF-GridNet approach.}
  \label{fig:seb-tp}
  \vspace{-0.2cm}
\end{figure*}

The 2D convolution layer in the encoder has a kernel size of (3,3) and a stride of 1, with input and output dimensions of 2 and 128, respectively. The attention block in the CMHA module consists of 1 layer, 4 parallel attention heads, and a 512-dimensional feed-forward network. The kernel size and stride of the torch.unfold function are both 1. In the full-band and sub-band modules, the BLSTM layer has 256 units. The cross-frame self-attention module uses 1 layer, 4 parallel attention heads, and a 512-dimensional feed-forward network. The number of TF-GridNet blocks is 6. The 2D transposed convolution layer in the TSE Decoder has the same kernel size and stride as the 2D convolution layer, with input and output dimensions of 256 and 2, respectively. The 2D transposed convolution layer in the PVAD Decoder has the same kernel size and stride as the 2D convolution layer, with input and output dimensions of 256 and 1, respectively. The kernel size and stride of the 1D convolution layer In the PVAD Decoder are 2 and 1, with input and output dimensions of 161 and 1, respectively. The kernel size and stride of the 1D transposed convolution layer are 2 and 1, with input and output dimensions both of 1. The hyperparameters for the USEF-TP model are summarized in Table~\ref{tab:hyp}. Additionally, we have implemented a Speaker-Embedding-Based Target speaker extraction and Personal voice activity detection model with the same backbone as USEF-TP, which we call SEB-TP. The architecture of the SEB-TP model is illustrated in Figure~\ref{fig:seb-tp}. We use ResNet34~\cite{10446780} to extract speaker embeddings in this work.

\subsection{Training Details}
For STFT, the window length is 20 ms, and the hop length is 10 ms. A 128-point Fourier transform is used to extract 161-dimensional complex STFT features at each frame. We trained our models using the Adam~\cite{KingBa15} optimizer, with an initial learning rate set to 1e-4. The learning rate was halved if the validation loss did not improve within 3 epochs. We set the weight of the multi-task loss \(\lambda_{1}\), \(\lambda_{2}\) = 1. The weights of the scene-aware TSE loss \(\alpha_{1}\) and \(\alpha_{2}\) are 1 and 0.01, respectively. In~\cite{pan2022usev}, the experimental results show relatively better performance when the energy loss weight for non-target speaker speech segments is set to 0.01. Moreover, when \(\alpha_{1} = 1\) and \(\alpha_{2} = 0.01\), the SI-SDR loss and energy loss are of a similar order of magnitude, with SI-SDR loss still being the predominant factor. Therefore, in this work, we adopt the same hyperparameter settings as~\cite{pan2022usev}, \(\alpha_{1} = 1\) and \(\alpha_{2} = 0.01\). No dynamic mixing or data augmentation was applied during training. The speech clips were truncated to 4 seconds during training, while the full speech clips were evaluated during inference. In the evaluation phase, each speaker in the mixed speech is considered as the target speaker in turn.

The model's training process is conducted in two stages. In the first stage, a scene-aware multi-task loss function is not applied; instead, the model is trained using SI-SDR and BCE as the primary loss functions. The scene-aware loss function is introduced during the second stage by incorporating energy, SI-SDR, and BCE losses from the first stage.  

\subsection{Evaluation Metrics}
We use SI-SDR or SI-SDR improvement (SI-SDRi) and signal-to-distortion ratio (SDR) or SDR improvement (SDRi) as objective metrics to assess the accuracy of TSE. We use the average energy (dB/s) to evaluate the model’s performance on TS speech segments.

We use the Recall (REC), Precision (PRE), F1 score (F1) and Accuracy (ACC) to evaluate the PVAD performance. We use Diarization Error Rate (DER), including Speaker Confusion (SC), False Alarm (FA), and Missed Detection (MS) to evaluate the SD performance. Collar tolerance is set to 0 seconds for LibriMix and SparseLibrimix, 0.25 seconds for CALLHOME dataset.

\section{Results and Discussions}
\label{sec:RD}
We present results from three sets of experiments. The first set is conducted on the LibriMix dataset. The second set focuses on the SparseLibriMix datasets. The third is performed on the CALLHOME dataset. In addition to presenting the TSE and PVAD results, we also integrate each speaker’s PVAD outputs to evaluate the model’s SD performance. Since our model is optimized only for PVAD and TSE, the SD results show greater variability across checkpoints. The main reason is that SD performance depends on the intermediate outputs of PVAD rather than being directly optimized by the loss function. To ensure a fair and stable evaluation, we mitigate checkpoint fluctuations for SD by averaging results from three nearby checkpoints around the best PVAD or TSE checkpoint.

\subsection{Results on the LibriMix Dataset}

In this section, we report the experimental results of the USEF-TP model on the LibriMix dataset. The LibriMix dataset includes both max and min modes. As the data overlap ratio in min mode is nearly 100\%, it is limitedly relevant to the purpose of this study. Therefore, we report only the results in max mode.
\begin{table}[]
\caption{Ablation study of the USEF-TP and SEB-TP model with the individual tasks on the LibriMix dataset for the max mode. ‘SEB-TP’ refers to the speaker-embedding-based model using the same backbone as USEF-TP.}
  \label{tab:abl_single_librimix}
\centering
\begin{tabular}{ccccc}
\Xhline{1px}
\multicolumn{5}{c}{\textbf{Personal Voice Activity Detection}}  \\ \Xhline{1px}
\multicolumn{1}{c|}{Model}     & ACC \(\uparrow\)        & F1 \(\uparrow\)        & PRE \(\uparrow\)        & Recall \(\uparrow\)       \\ \hline
\multicolumn{1}{c|}{SEB-TP}   & 0.900       &  0.921       &  0.972            &  0.874               \\
\multicolumn{1}{c|}{\quad \quad \quad  - Only PVAD} &  0.877    &  0.905       &  0.936           &   0.875            \\
\multicolumn{1}{c|}{USEF-TP}   &  0.968       &  0.977       &  0.983            &  0.970               \\
\multicolumn{1}{c|}{\quad \quad \quad - Only PVAD} &  0.956    &  0.967       &  0.964            &   0.970             \\ \Xhline{1px}
\multicolumn{5}{c}{\textbf{Speaker Diarization}}                                                        \\ \Xhline{1px}
\multicolumn{1}{c|}{Model}     & DER (\%) \(\downarrow\)        & MI (\%) \(\downarrow\)       & FA (\%) \(\downarrow\)        & CF (\%) \(\downarrow\)          \\ \hline
\multicolumn{1}{c|}{ SEB-TP}   &  15.82       &  13.87       &  1.89            &  0.06               \\
\multicolumn{1}{c|}{\quad \quad \quad  - Only PVAD} &  16.78    &  12.39       &  4.14            &   0.25             \\
\multicolumn{1}{c|}{USEF-TP}   &  4.92        &  3.91       &  0.99        &  0.02               \\
\multicolumn{1}{c|}{\quad \quad \quad - Only PVAD} &  6.48        &  4.21       &  2.20        &  0.06               \\ \Xhline{1px}
\multicolumn{5}{c}{\textbf{Target Speaker Extraction}}                                                        \\ \Xhline{1px}
\multicolumn{1}{c|}{Model}     & \multicolumn{2}{c}{SDRi (dB) \(\uparrow\)} & \multicolumn{2}{c}{SI-SDRi (dB) \(\uparrow\)} \\ \hline
\multicolumn{1}{c|}{ SEB-TP}   & \multicolumn{2}{c}{ 12.38}  & \multicolumn{2}{c}{ 12.07}                    \\
\multicolumn{1}{c|}{\quad \quad \quad  - Only TSE} & \multicolumn{2}{c}{ 12.23}  & \multicolumn{2}{c}{ 11.80}                   \\
\multicolumn{1}{c|}{USEF-TP}   & \multicolumn{2}{c}{17.38}     & \multicolumn{2}{c}{16.82}        \\
\multicolumn{1}{c|}{\quad \quad \quad - Only TSE} & \multicolumn{2}{c}{17.11}     & \multicolumn{2}{c}{16.51}        \\ \Xhline{1px}
\end{tabular}
\end{table}
\subsubsection{Comparison with Speaker Embedding based and Single Task Models}
\label{subsubsec: single-task}
Table~\ref{tab:abl_single_librimix} shows the comparison results of the USEF-TP and SEB-TP models, which use speaker embedding and the same backbone as USEF-TP. The results in Table~\ref{tab:abl_single_librimix} show that USEF-TP outperforms SEB-TP in all metrics for the PVAD and SD tasks, with notable improvements in Recall (0.970 vs. 0.874) and DER (4.92\% vs 15.82 \%). It indicates that the speaker-embedding-free framework in USEF-TP helps improve the model’s performance in the PVAD task.

To demonstrate the performance of USEF-TP and SEB-TP on the single tasks of TSE and PVAD, we remove the TSE and PVAD part, respectively. Table~\ref{tab:abl_single_librimix} shows the ablation study results of the USEF-TP and SEB-TP model with the single tasks on the LibriMix dataset. In the PVAD task, aside from Recall, the USEF-TP model outperforms the single PVAD model in terms of ACC, F1, and PRE. By aggregating the PVAD predictions for each speaker in the mixed speech, we can obtain the corresponding SD results. The USEF-TP model significantly outperforms the single PVAD model in the SD task, particularly regarding DER and MI (DER: 4.92\% vs. 6.48\% and MI: 3.91\% vs. 4.21\%). The results of the joint model and single-task model on the PVAD and SD tasks suggest that joint task training may enhance the model’s PVAD performance.

For the TSE task, compared to SEB-TP, USEF-TP achieves relative improvements of 40.4\% in SDRi (17.38 dB vs. 12.38 dB) and 39.4\% in SI-SDRi (16.82 dB vs. 12.07 dB). The results suggest that the speaker-embedding-free framework may enhance the model’s TSE performance. The USEF-TP model's SDRi and SI-SDRi are 0.27 dB (17.38 dB vs. 17.11 dB) and 0.31 dB (16.82 dB vs. 16.51 dB) higher than those of the single-task model, respectively. It indicates that joint training may enhance the model’s TSE performance.

Overall, the USEF-TP model outperforms the SEB-TP and single-task models (Only PVAD or Only TSE) across all tasks. It indicates that speaker-embedding-free framework and joint task training can enhance performance on each subtask, particularly in the SD task.

 
%
\begin{table}[]
\vspace{-2cm}
\caption{Ablation study of the USEF-TP model using the Scene-aware Loss (SL) function with or without the Interaction Module (IM) on the LibriMix dataset for max mode. 'M' denotes the mixed speech. 'IM' and 'SL' denote the Interaction Module and Scene-aware Loss function, respectively. 'TA' and 'TS' denote the target speaker active and target speaker silent speech clips, respectively.}
  \label{tab:abl_imsl_librimix}
\centering
\begin{tabular}{ccccccc}
\Xhline{1px}
\multicolumn{7}{c}{\textbf{Personal Voice Activity Detection}}                                                                                                                                               \\ \Xhline{1px}
\multicolumn{1}{c|}{Exp}                  & IM                  & \multicolumn{1}{c|}{SL}                  & ACC \(\uparrow\)  & F1 \(\uparrow\)                          & Precision \(\uparrow\) & Recall \(\uparrow\) \\ \hline
\multicolumn{1}{c|}{1}                    & \XSolidBrush                   & \multicolumn{1}{c|}{\XSolidBrush}                   & 0.954  &  0.964                         &   0.964     &  0.946       \\
\multicolumn{1}{c|}{2}                    &  \XSolidBrush                   & \multicolumn{1}{c|}{ \CheckmarkBold}                   &  0.965  &   0.974                        &  0.985      &  0.962        \\
\multicolumn{1}{c|}{ 3}                    & \CheckmarkBold                   & \multicolumn{1}{c|}{\XSolidBrush}                   &  0.969  &   0.976                        &  0.984      &  0.966        \\
\multicolumn{1}{c|}{ 4}                    & \CheckmarkBold                    & \multicolumn{1}{c|}{\CheckmarkBold }                   &  0.970  &   0.977                        &  0.983      &  0.970    \\ \Xhline{1px}
\multicolumn{7}{c}{\textbf{Speaker Diarization}}                                                                                                                                                 \\ \Xhline{1px}
\multicolumn{1}{c|}{Exp}                  & IM                  & \multicolumn{1}{c|}{SL}                  & DER (\%) \(\downarrow\)  & MI (\%) \(\downarrow\)                         & FA (\%) \(\downarrow\)      & CF (\%) \(\downarrow\)   \\ \hline
\multicolumn{1}{c|}{ 5}                    & \XSolidBrush                   & \multicolumn{1}{c|}{\XSolidBrush}                   &  5.65  &  4.30                         &  1.33      &  0.02       \\
\multicolumn{1}{c|}{6}                    &  \XSolidBrush                   & \multicolumn{1}{c|}{ \CheckmarkBold}                   &  5.01  &   4.00                        &  0.99      &  0.02        \\
\multicolumn{1}{c|}{7}                    & \CheckmarkBold                   & \multicolumn{1}{c|}{\XSolidBrush}                   &  5.38  &  4.46                         &  0.90      &  0.02       \\
\multicolumn{1}{c|}{8}                    & \CheckmarkBold                   & \multicolumn{1}{c|}{\CheckmarkBold}                   &  4.92  & 3.91                         &  0.99      &  0.02       \\ \Xhline{1px}
\multicolumn{7}{c}{\textbf{Target Speaker Extraction}}                                                                                                                                                \\ \Xhline{1px}
\multicolumn{1}{c|}{\multirow{2}{*}{Exp}} & \multirow{2}{*}{IM} & \multicolumn{1}{c|}{\multirow{2}{*}{SL}} & \multicolumn{2}{c|}{Overall}         & TA        & TS     \\
\multicolumn{1}{c|}{}                     &                     & \multicolumn{1}{c|}{}                    & SDRi (dB) \(\uparrow\) & \multicolumn{1}{c|}{SI-SDRi (dB) \(\uparrow\)} & SI-SDRi (dB) \(\uparrow\)  & Power (dB/s) \(\downarrow\) \\ \hline
\multicolumn{1}{c|}{M}                & -                   & \multicolumn{1}{c|}{-}                   & -     & \multicolumn{1}{c|}{-}       & -         & 8.47  \\ \hline
\multicolumn{1}{c|}{ 9}                    & \XSolidBrush                   & \multicolumn{1}{c|}{\XSolidBrush}                   & 17.13 & \multicolumn{1}{c|}{16.61}   & 15.15     & -7.38  \\
\multicolumn{1}{c|}{ 10}                    &  \XSolidBrush                   & \multicolumn{1}{c|}{ \CheckmarkBold}                   &  17.34  &  \multicolumn{1}{c|}{ 16.67}                        &  15.37      &  -15.68        \\
\multicolumn{1}{c|}{ 11}                    & \CheckmarkBold                   & \multicolumn{1}{c|}{\XSolidBrush}                   & 17.16 & \multicolumn{1}{c|}{16.57}   & 15.35     & -9.73 \\
\multicolumn{1}{c|}{ 12}                    & \CheckmarkBold                   & \multicolumn{1}{c|}{\CheckmarkBold}                   & 17.38 & \multicolumn{1}{c|}{16.82}   & 15.51     & -17.11 \\ \Xhline{1px}
\end{tabular}
\end{table}
\subsubsection{Ablation Study for Interaction Module and Loss Function}

Table~\ref{tab:abl_imsl_librimix} presents the ablation study results for the USEF-TP model with and without the Interaction Module (IM) and Scene-aware Loss function (SL). The experiments are conducted in max mode on the LibriMix dataset. The table is divided into three sections, showing the model’s performance on the PVAD, SD, and TSE tasks. When the SL or IM is introduced (Exp 2, 3), the model’s ACC, F1, and Recall all improve (ACC: 0.954 \(\rightarrow\)0.965, 0.969, F1: 0.964 \(\rightarrow\) 0.974, 0.976; Recall: 0.946 \(\rightarrow\) 0.962, 0.966).  Moreover, when the SL and IM are both introduced (Exp 4), USEF-TP achieves the best performance. For the SD task, after introducing the  SL (Exp 6) or IM (Exp 7), the model's DER decrease ( DER: 5.65\% \(\rightarrow\) 5.01\%, 5.38\%). With the addition of both SL and IM (Exp 8), the model's DER and MI show a further decrease(DER: 5.65\% \(\rightarrow\) 4.92\%, MI: 4.30\% \(\rightarrow\) 3.91\%). The results indicate that the SL and IM provides some benefits for the PVAD task. 

For the TSE task, When the SL is introduced (Exp 10), both the SI-SDRi of the entire audio and the TA (target speaker activate) segment increase (16.61 dB \(\rightarrow\) 16.67 dB, 15.13 dB \(\rightarrow\) 15.37 dB). Moreover, the power of the TS (target speaker silent) segments decreases significantly. This indicates that SL improves the model’s ability to handle both TA and TS segments. When the IM is introduced (Exp 11), the overall SI-SDRi show a slight decrease (16.61 dB \(\rightarrow\) 16.57 dB). In the TA conditions, the model’s SI-SDRi improved by 0.2 dB (15.15 dB \(\rightarrow\) 15.51 dB). Under the TS conditions, the decrease in power of the speech segments is not significant. It suggests that the IM mainly enhances performance in overlapped speech segments in the TSE task, with limited improvement in handling TS segments. With joint use of SL and IM, the model’s SI-SDRi improved by 0.25 dB overall and by 0.16 dB in TA, with average energy decreasing from the original -9.73 dB/s to -17.11 dB/s. These results indicate that using SL and IM can enhance the model’s overall TSE performance, particularly with a notable improvement in TS segments by reducing the inference speakers' artifacts.
\begin{table}[]
\caption{Comparison with the previous models on different tasks on the LibriMix dataset for max mode.'\(\dagger\)' indicates the model implemented by ourselves.}
\label{tab:cmp_librimix}
\centering
\begin{tabular}{cccccc}
\Xhline{1px}
\multicolumn{6}{c}{\textbf{Personal Voice Activity Detection}}                                                                                                         \\ \Xhline{1px}
\multicolumn{1}{c|}{Model}    & \multicolumn{1}{c|}{ACC  \(\uparrow\)} & \multicolumn{1}{c|}{F1  \(\uparrow\)} & \multicolumn{1}{c|}{PRE  \(\uparrow\)} & \multicolumn{1}{c|}{Recall \(\uparrow\)} & Parameter (M) \\ \hline
\multicolumn{1}{c|}{PVAD 1.0\(\dagger\)~\cite{Ding2020}} & \multicolumn{1}{c|}{0.816} & \multicolumn{1}{c|}{ 0.856} & \multicolumn{1}{c|}{ 0.893} & \multicolumn{1}{c|}{ 0.822} & 2.3 \\
\multicolumn{1}{c|}{PVAD 2.0\(\dagger\)~\cite{Ding2022PersonalV2}} & \multicolumn{1}{c|}{0.844} & \multicolumn{1}{c|}{ 0.880} & \multicolumn{1}{c|}{0.906} & \multicolumn{1}{c|}{ 0.854} & 10.4 \\
\multicolumn{1}{c|}{SEB-TP}  & \multicolumn{1}{c|}{ 0.900} & \multicolumn{1}{c|}{0.921} & \multicolumn{1}{c|}{ 0.972} & \multicolumn{1}{c|}{ 0.874} & 15.1  \\
\multicolumn{1}{c|}{USEF-TP}  & \multicolumn{1}{c|}{\textbf{0.970}} & \multicolumn{1}{c|}{\textbf{0.977}} & \multicolumn{1}{c|}{\textbf{0.983}} & \multicolumn{1}{c|}{\textbf{0.970}} & 15.1 \\ \Xhline{1px}
\multicolumn{6}{c}{\textbf{Speaker Diarization}}                                                                                                                        \\ \Xhline{1px}
\multicolumn{1}{c|}{Model}    & \multicolumn{1}{c|}{DER (\%) \(\downarrow\)} & \multicolumn{1}{c|}{MI (\%)  \(\downarrow\)} & \multicolumn{1}{c|}{FA (\%) \(\downarrow\)} & \multicolumn{1}{c|}{CF (\%) \(\downarrow\)} & Parameter (M) \\ \hline
\multicolumn{1}{c|}{ PVAD 1.0\(\dagger\)~\cite{Ding2020}} & \multicolumn{1}{c|}{27.69} & \multicolumn{1}{c|}{ 22.07} & \multicolumn{1}{c|}{4.88} & \multicolumn{1}{c|}{0.74} &  2.3 \\
\multicolumn{1}{c|}{PVAD 2.0\(\dagger\)~\cite{Ding2022PersonalV2}} & \multicolumn{1}{c|}{22.16} & \multicolumn{1}{c|}{ 16.16} & \multicolumn{1}{c|}{5.51} & \multicolumn{1}{c|}{0.49} &  10.4 \\
\multicolumn{1}{c|}{SEB-TP}   & \multicolumn{1}{c|}{ 15.82} & \multicolumn{1}{c|}{13.87} & \multicolumn{1}{c|}{1.89} & \multicolumn{1}{c|}{ 0.06} & 15.1 \\
\multicolumn{1}{c|}{TS-VAD~\cite{medennikov2020target}}   & \multicolumn{1}{c|}{7.28} & \multicolumn{1}{c|}{3.61} & \multicolumn{1}{c|}{2.78} & \multicolumn{1}{c|}{0.89} &  39.50 \\
\multicolumn{1}{c|}{USED-F~\cite{Ao_2024}}   & \multicolumn{1}{c|}{\textbf{4.75}} & \multicolumn{1}{c|}{\textbf{2.18}} & \multicolumn{1}{c|}{2.16} & \multicolumn{1}{c|}{0.42} &  23.12 \\
\multicolumn{1}{c|}{USEF-TP}  & \multicolumn{1}{c|}{ 4.92} & \multicolumn{1}{c|}{ 3.91} & \multicolumn{1}{c|}{\textbf{0.99}} & \multicolumn{1}{c|}{\textbf{ 0.02}} & 15.1 \\ \Xhline{1px}
\multicolumn{6}{c}{\textbf{Target Speaker Extraction}}                                                                                                                   \\ \Xhline{1px}
\multicolumn{1}{c|}{Model}    & \multicolumn{2}{c|}{SDRi (dB) \(\uparrow\)} & \multicolumn{2}{c|}{SI-SDRi (dB) \(\uparrow\)} & Parameter (M) \\ \hline
\multicolumn{1}{c|}{SpEx+~\cite{ge2020spex+}}    & \multicolumn{2}{c|}{10.11} & \multicolumn{2}{c|}{9.06} &  16.35 \\
\multicolumn{1}{c|}{SEB-TP}   & \multicolumn{2}{c|}{ 12.38} & \multicolumn{2}{c|}{12.07} &  15.1 \\
\multicolumn{1}{c|}{USED-F~\cite{Ao_2024}}   & \multicolumn{2}{c|}{13.22} & \multicolumn{2}{c|}{12.70} &  23.12 \\
\multicolumn{1}{c|}{USEF-TP}  & \multicolumn{2}{c|}{\textbf{17.38}} & \multicolumn{2}{c|}{\textbf{16.82}} & 15.1 \\ \Xhline{1px}
\end{tabular}
\end{table}
\begin{table}[]
\vspace{-1cm}
\caption{Ablation study of the USEF-TP and SEB-TP model with the individual tasks on the SparseLibriMix dataset. Due to space constraints, we present only the results for overlap ratios of 0\% / 20\% / 40\% in the table. ‘SEB-TP’ refers to the speaker-embedding-based model using the same backbone as USEF-TP.}
  \label{tab:abl_single_sparselibrimix}
\centering
\resizebox{\linewidth}{!}{\LARGE
\begin{tabular}{ccccc}
\Xhline{1.5px}
\multicolumn{5}{c}{\makecell{\textbf{Personal Voice Activity Detection} \\ 
(with overlap ratios of 0\% / 20\% / 40\%)}}                                                                                                                                           \\ \Xhline{1.5px}
\multicolumn{1}{c|}{Model}        & ACC \(\uparrow\)                                   & F1  \(\uparrow\)                          & Precision \(\uparrow\)                     & Recall \(\uparrow\)                       \\ \hline
\multicolumn{1}{c|}{SEB-TP}      &  0.896 / 0.886 /  0.888         & 0.849 /  0.880 / 0.891 &  0.891 /  0.932 /  0.942  &  0.810 /  0.834 /  0.845  \\
\multicolumn{1}{c|}{\quad \quad \quad  - Only PVAD} &  0.763 / 0.810 /  0.827          &  0.711 /  0.822 /  0.848 &  0.633 /  0.771 /  0.808  &  0.809 /  0.881 /  0.892 \\
\multicolumn{1}{c|}{USEF-TP}      &   0.969 /  0.971 /  0.971         &  0.956 /  0.971 /  0.973 &  0.951 /  0.971 /  0.971  &  0.962 /  0.971 /  0.976  \\
\multicolumn{1}{c|}{\quad \quad \quad - Only PVAD} &  0.922 /  0.928 /  0.933          &  0.896 /  0.930 /  0.939 &  0.861 /  0.906 /  0.922  &  0.935 /  0.954 /  0.957 \\ \Xhline{1.5px}
\multicolumn{5}{c}{\makecell{\textbf{Speaker Diarization} \\
(with overlap ratios of 0\% / 20\% / 40\%)}}                                                                                                                                             \\ \Xhline{1.5px}
\multicolumn{1}{c|}{Model}        & DER (\%) \(\downarrow\)                                  & MI (\%)  \(\downarrow\)                         & FA (\%)  \(\downarrow\)                          & CF (\%) \(\downarrow\)                          \\ \hline
\multicolumn{1}{c|}{ SEB-TP}      &   28.64 /  23.63 /  22.13         &  20.60 /  18.57 /  17.90 &  6.66 /  4.42 /  3.76  &  1.38 /  0.65 /  0.46  \\
\multicolumn{1}{c|}{\quad \quad \quad  - Only PVAD} &  53.08 /  33.95 /  28.95          &  21.38 /  16.75 /  15.80 &  25.40 /  14.81 /  11.50  &  6.30 /  2.38 /  1.65 \\
\multicolumn{1}{c|}{USEF-TP}      & 8.33 / 5.62 /  5.23         &  3.54 /  2.87 /  2.71 &  4.57 /  2.64 /  2.46  &  0.22 /  0.12 /  0.06 \\
\multicolumn{1}{c|}{\quad \quad \quad - Only PVAD} &  21.22 /  14.00 /  11.74      &  3.86 /  3.29 /  3.27 &  16.40 /  10.33 /  8.22 &  0.96 /  0.38 /  0.24\\ \Xhline{1.5px}
\multicolumn{5}{c}{\makecell{\textbf{Target Speaker Extraction} \\
(with overlap ratios of 0\% / 20\% / 40\%)}}                                                                                                                                            \\ \Xhline{1.5px}
\multicolumn{1}{c|}{Model}        & \multicolumn{2}{c}{SDRi (dB) \(\uparrow\)}                                               & \multicolumn{2}{c}{SI-SDRi (dB) \(\uparrow\)}                                    \\ \hline
\multicolumn{1}{c|}{ SEB-TP}      &   \multicolumn{2}{c}{ 15.51 /  14.41 /  14.46}        &  \multicolumn{2}{c}{ 14.79 /  13.19 /  13.16}\\
\multicolumn{1}{c|}{\quad \quad \quad  - Only TSE} &   \multicolumn{2}{c}{ 14.73 /  14.11 /  13.87}         &  \multicolumn{2}{c}{ 13.26 /  12.41 /  12.21}\\
\multicolumn{1}{c|}{USEF-TP}      & \multicolumn{2}{c}{20.39 / 19.27 / 19.28}                & \multicolumn{2}{c}{19.78 / 18.68 / 18.68}        \\
\multicolumn{1}{c|}{\quad \quad \quad - Only TSE} & \multicolumn{2}{c}{19.62 / 18.70 / 18.52}                & \multicolumn{2}{c}{19.05 / 18.11 / 17.92}        \\ \Xhline{1.5px}
\end{tabular}
}
\end{table}

\subsubsection{Comparison With Previous Models}
Table~\ref{tab:cmp_librimix} compares the performance of the USEF-TP model across three tasks (PVAD, SD, and TSE). The USEF-TP model is evaluated against multiple baseline models for each task. For the PVAD experimental results, USEF-TP significantly outperforms PVAD 1.0 and PVAD 2.0 in ACC, F1, PRE, and Recall. Specifically, USEF-TP achieves a relative improvement of 14.9 \% (0.970 vs. 0.844) in ACC and 13.6 \% (0.970 vs. 0.854) in Recall compared to PVAD 2.0. It demonstrates that USEF-TP has a clear advantage over PVAD 1.0 and PVAD 2.0 models. Similarly, we aggregate the PVAD results for each speaker in the mixed speech within the test dataset to compute the model’s SD results and compare them with other SD baseline models. The experimental results show that USEF-TP outperforms other models in DER, FA, and CF metrics, with a particularly notable result in CF (0.02 \%). It indicates that the USEF-TP model has lower error and confusion rates in the SD task. Although DER and MI is slightly higher than USED-F ( DER:4.92 \% vs. 4.75 \%; MI: 3.91 \% vs. 2.18 \%), the overall performance remains superior to other models.

For the TSE results, USEF-TP significantly surpasses other models on SDRi and SI-SDRi metrics, indicating a more substantial improvement in the TSE task. Specifically, compared to the USED-F model, the USEF-TP model achieves improvements of 31.5 \% (17.38 dB vs. 13.22 dB) in SDRi and 32.4 \% (16.82 dB vs. 12.70 dB) in SI-SDRi. It demonstrates that the USEF-TP model excels in extracting high-quality target speaker signals

The USEF-TP model outperforms other baseline models across all tasks, with a notable advantage in the SDRi and SI-SDRi for the TSE task. Additionally, the USEF-TP model demonstrates significant improvement in all metrics for PVAD, indicating that the joint tasks training enhances the model's performance on both TSE and PVAD tasks.

\subsection{Results on the SparseLibriMix Dataset}
In this section, we report the experimental results of the USEF-TP model on the SparseLibriMix dataset, which includes multi-speaker mixed test speech with six different overlap ratios: 0\%, 20\%, 40\%, 60\%, 80\%, and 100\%. The primary goal of this set of experiments is to evaluate the proposed model's performance under sparse overlap conditions. The demo are available at this link \footnotemark[3]\footnotetext[3]{\url{https://github.com/ZBang/USEF-TP}}.

\subsubsection{Comparison with Speaker Embedding based and Single Task Models}
Table~\ref{tab:abl_single_sparselibrimix} presents the {comparison} results on the SparseLibriMix dataset, comparing the USEF-TP model with single-task models containing USEF-TP and SEB-TP (Only PVAD) or USEF-TP and SEB-TP (Only TSE) across different overlap ratios (0\%, 20\%, and 40\%). The USEF-TP and SEB-TP models outperform the Only PVAD models in ACC, F1, Precision, and Recall across all overlap ratios, indicating that the joint model USEF-TP achieves better overall performance in the PVAD task, with notably stable performance at higher overlap ratios (40\%). Additionally, the USEF-TP model achieves significantly lower DER, MI, FA, and CF scores across all overlap ratios, especially in DER at an overlap ratio of 0\% (DER: 8.33\% vs. 21.22\%). It suggests that multi-task joint training can significantly enhance PVAD performance under sparse overlap conditions. By comparing the experimental results of USEF-TP and SEB-TP, we observe that USEF-TP achieves significant improvements in both F1 score and DER across different overlap ratios, with a notable improvement in Recall (overlap ration 0\% : 0.962 vs. 0.810). This demonstrates that adopting the speaker-embedding-free framework enhances the model’s performance in the PVAD task.

The USEF-TP model achieves higher SDRi and SI-SDRi scores than the Only TSE model, with the largest gap observed at 0\% overlap (SDRi: 20.39 dB vs. 19.62 dB, SI-SDRi: 19.78 dB vs. 19.05 dB). It indicates that under sparse overlap conditions, multi-task joint training can significantly enhance TSE performance. Furthermore, the joint model exhibits more stable performance across different overlap ratios compared to the only TSE model. Additionally, when the overlap ratio is 0, USEF-TP achieves a 33.7\% (19.78 dB vs. 14.79 dB) improvement in SI-SDRi compared to SEB-TP. These results indicate that the speaker-embedding-free framework enhances the model’s performance in the TSE task. Besides speaker identity, other information in the reference speech, such as contextual cues, may also contribute to extracting the target speaker’s clean speech.

From the experimental results in Table~\ref{tab:abl_single_sparselibrimix}, we can conclude that the USEF-TP model outperforms the SEB-TP model on both tasks. The results demonstrate the effectiveness of the speaker-embedding-free framework in handling mixtures with different overlap ratios. Moreover the USEF-TP model outperforms the single-task models across all tasks, maintaining strong performance at low and high overlap ratios. It indicates that joint task training enhances the model’s performance and improves its stability.
\begin{table}[]
\vspace{-1.8cm}
\caption{Ablation study of the USEF-TP model using the scene-aware loss function with or without the Interaction module on the SparseLibriMix dataset. 'M' denotes the mixed speech. 'IM' and 'SL' denote the Interaction module and scene-aware loss function, respectively. 'TA' and 'TS' denote the target speaker active and target speaker silent speech clips, respectively. Due to space constraints, we present only the results for overlap ratios of 0\% / 20\% / 40\% in the table.}
  \label{tab:abl_imsl_sparselibrimix}
\centering
\resizebox{\textwidth}{!}{\Huge
\begin{tabular}{ccccccc}
\Xhline{2px}
\multicolumn{7}{c}{\makecell{\textbf{Personal Voice Activity Detection} \\ 
(with overlap ratios of 0\% / 20\% / 40\%)}}                                                                                                                                                                                                           \\ \Xhline{2px}
\multicolumn{1}{c|}{Exp}                  & IM                  & \multicolumn{1}{c|}{SL}                  & ACC \(\uparrow\)                  & F1 \(\uparrow\)                                        & Precision \(\uparrow\)             & Recall  \(\uparrow\)                 \\ \hline
\multicolumn{1}{c|}{1}                    & \XSolidBrush                   & \multicolumn{1}{c|}{\XSolidBrush}                   &  0.962 /  0.963 /  0.964    &  0.946 /  0.962 /  0.967                         &  0.949 /  0.963 /  0.966    &  0.943 /  0.962 /  0.968       \\
\multicolumn{1}{c|}{ 2}                   &  \XSolidBrush                  & \multicolumn{1}{c|}{ \CheckmarkBold}                   &  0.967 /  0.970 /  0.970    &  0.955 /  0.965 /  0.972                         &  0.945 /  0.965 /  0.969    &  0.965 /  0.975 /  0.976       \\
\multicolumn{1}{c|}{ 3}                    & \CheckmarkBold                   & \multicolumn{1}{c|}{\XSolidBrush}                   &  0.967 /  0.966 /  0.965    &  0.954 /  0.966 /  0.965                         &  0.951 /  0.967 /  0.967    &  0.958 /  0.965 /  0.970       \\
\multicolumn{1}{c|}{ 4}                    & \CheckmarkBold                   & \multicolumn{1}{c|}{\CheckmarkBold}                   &  0.969 /  0.971 /  0.971         &  0.956 /  0.971 /  0.973 &  0.951 /  0.971 /  0.971  &  0.962 /  0.971 /  0.976       \\ \Xhline{2px}
\multicolumn{7}{c}{\makecell{\textbf{Speaker Diarization} \\
(with overlap ratios of 0\% / 20\% / 40\%)}}                                                                                                                                                                                                             \\ \Xhline{2px}
\multicolumn{1}{c|}{Exp}                  & IM                  & \multicolumn{1}{c|}{SL}                  & DER (\%) \(\downarrow\)                 & MI (\%) \(\downarrow\)                                       & FA  (\%) \(\downarrow\)                 & CF (\%) \(\downarrow\)                     \\ \hline
\multicolumn{1}{c|}{ 5}                    & \XSolidBrush                   & \multicolumn{1}{c|}{\XSolidBrush}                   &  9.53 /  6.34 /  5.73    &  3.88 /  2.95 /  2.74                         &  5.34 /  3.25 /  2.92    &  0.31 /  0.14 /  0.07       \\
\multicolumn{1}{c|}{ 6}                   &  \XSolidBrush                  & \multicolumn{1}{c|}{ \CheckmarkBold}                   &  9.06 /  5.98 /  5.36    &  3.66 /  2.70 /  2.54                         &  5.09 /  3.17 /  2.76    &  0.31 /  0.11 /  0.07       \\
\multicolumn{1}{c|}{ 7}                    & \CheckmarkBold                   & \multicolumn{1}{c|}{\XSolidBrush}                   &  9.21 /  6.22 /  5.65    &  3.82 /  3.21 /  2.94                         &  5.07 /  2.87 /  2.64    &  1.22 /  0.14 /  0.08       \\
\multicolumn{1}{c|}{ 8}                    & \CheckmarkBold                   & \multicolumn{1}{c|}{\CheckmarkBold}                   &   8.33 /  5.62 /  5.23         &  3.54 /  2.87 /  2.71 &  4.57 /  2.64 /  2.46  &  0.22 /  0.12 /  0.06       \\ \Xhline{2px}
\multicolumn{7}{c}{\makecell{\textbf{Target Speaker Extraction} \\
(with overlap ratios of 0\% / 20\% / 40\%)}}                                                                                                                                                                                                            \\ \Xhline{2px}
\multicolumn{1}{c|}{\multirow{2}{*}{Exp}} & \multirow{2}{*}{IM} & \multicolumn{1}{c|}{\multirow{2}{*}{SL}} & \multicolumn{2}{c|}{Overall}                                       & TA                    & TS                       \\
\multicolumn{1}{c|}{}                     &                     & \multicolumn{1}{c|}{}                    & SDRi (dB) \(\uparrow\)                 & \multicolumn{1}{c|}{SI-SDRi (dB) \(\uparrow\)}               & SI-SDRi (dB) \(\uparrow\)               & Power (dB/s) \(\downarrow\)                    \\ \hline
\multicolumn{1}{c|}{M}                & -                   & \multicolumn{1}{c|}{-}                   & -                     & \multicolumn{1}{c|}{-}                     & -                     & 3.73 / 5.94 / 6.81     \\ \hline
\multicolumn{1}{c|}{ 9}                    & \XSolidBrush                   & \multicolumn{1}{c|}{\XSolidBrush}                   & 20.03 / 18.92 / 18.63  & \multicolumn{1}{c|}{19.37 / 18.29 / 17.96}& 11.62 / 14.36 / 14.83& -1.14 / -2.70 / -3.89 \\
\multicolumn{1}{c|}{ 10}                    &  \XSolidBrush    & \multicolumn{1}{c|}{ \CheckmarkBold}   &  20.15 /  19.11 /  18.86   & \multicolumn{1}{c|}{ 19.43 /  18.41 /  18.19}                         &  11.63 /  14.45 /  14.91    &  -2.60 /  -5.61 /  -7.61       \\
\multicolumn{1}{c|}{ 11}                    & \CheckmarkBold                   & \multicolumn{1}{c|}{\XSolidBrush}                   & 20.38 / 18.96 / 18.62 & \multicolumn{1}{c|}{19.76 / 18.34 / 17.99} & 11.92 / 14.41 / 14.88 & -1.79 / -3.71 / -5.07 \\
\multicolumn{1}{c|}{ 12}                    & \CheckmarkBold                   & \multicolumn{1}{c|}{\CheckmarkBold}                   & 20.39 / 19.27 / 19.28  & \multicolumn{1}{c|}{19.78 / 18.68 / 18.68} & 11.90 / 14.65 / 15.14 & -3.35 / -6.79 / -8.89 \\ \Xhline{2px}
\end{tabular}
}
\end{table}
\subsubsection{Ablation Study for Interaction Module and Loss Function}
Table~\ref{tab:abl_imsl_sparselibrimix} presents the ablation study results on the impact of including the IM and SL on the performance of the USEF-TP model. It examines the model’s performance on the SparseLibriMix dataset under different combinations of these two modules. For the PVAD task, incorporating the SL (Exp 3) or IM (Exp  3) leads to improvements in ACC, F1, Precision, and Recall, with further enhancements observed when SL and IM (Exp 4) are both applied. Furthermore, comparing EXP 5,6,7,8 shows that at a 0\% overlap ratio, the addition of SL and IM leads to an apparent decrease in DER ( 9.56\% \(\rightarrow\) 9.06\%, 9.21\% \(\rightarrow\) 8.33\% ). These results indicate that the IM and SL can enhance the model’s PVAD performance even under sparse overlap conditions.

For the TSE task, as SL and IM are added (from Exp 9 to Exp 12), both SDRi and SI-SDRi improve incrementally. It indicates that the IM and SL positively enhance overall TSE performance. Comparing Exp 9 to Exp 12 reveals that as the overlap ratio increases, overall SI-SDRi tends to decrease, while SI-SDRi under TA conditions shows an upward trend. It suggests the model’s capability to handle TS speech segments diminishes with higher overlap ratios. However, Exp 12 maintains a more robust performance in overall SI-SDRi, and compared to Exp 11, Exp 10 and 12 show a more noticeable reduction in average energy under TS conditions. It indicates that SL effectively reduces the power of interfering speakers or noise in silent segments.

In summary, under various overlap conditions, the joint model USEF-TP with both IM and SL outperforms versions without these modules or with only a single module across all tasks, demonstrating the effectiveness of IM and SL in enhancing model performance and stability.
\begin{figure*}[t]
  \centering
  \vspace{-10pt}
  \includegraphics[width=0.85\linewidth]{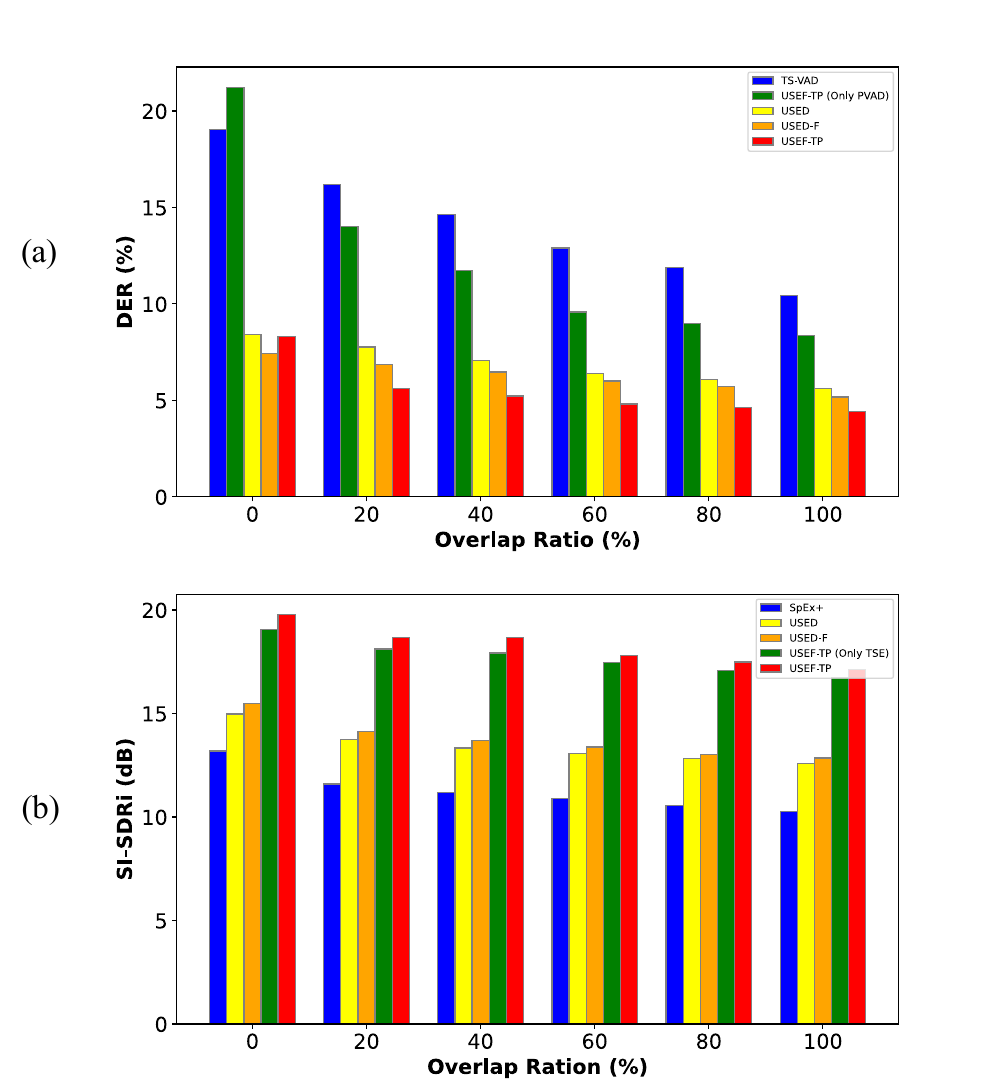}
  \caption{Comparison with previous models on the SparseLibriMix for overlap ratio from 0\% to  100\%. Due to space constraints, we just present the results of SD and TSE tasks. (a) is the result of the SD task. (b) is the result of the TSE task. The results of the TS-VAD, SpEx+, USED, and USEF-F are given in~\cite{Ao_2024}.}
  \label{fig:cmp_sparselibrimix}
  \vspace{-0.3cm}
\end{figure*}
\subsubsection{Comparison With Previous Models}
Figure~\ref{fig:cmp_sparselibrimix} illustrates the performance of several models on the SD and TSE tasks on the SparseLibriMix dataset across different overlap ratios (0\% to 100\%). It includes four models: TS-VAD~\cite{medennikov2020target}, SpEx+~\cite{ge2020spex+}, USED-F~\cite{Ao_2024}, USEF-TP, as well as various task settings for USEF-TP. The plot (a) shows the DER for the SD task, while the plot (b) displays the SI-SDRi for the TSE task. From (a), we can see that the DER of the models generally increases as the overlap ratio decreases, indicating that as the overlapping segments in the audio decrease, the SD task becomes more challenging. Across different overlap ratios, USEF-TP consistently achieves a lower DER than the Only PVAD model, indicating that the joint model design helps reduce DER and improve SD performance. Except at the 0\% overlap ratio, where USEF-TP’s DER is slightly higher than that of the USED-F model, USEF-TP outperforms the other models in DER at all other overlap ratios. This advantage is particularly notable at low overlap rates (20\% and 40\%). The USEF-TP model maintains a low DER across various overlap ratios, demonstrating strong performance and stability.

For the TSE task, as shown in (b), the SI-SDRi of the models generally decreases as the overlap ratio increases, indicating that higher overlap makes the TSE task more challenging. The SI-SDRi of USEF-TP is significantly higher than that of the single-task model (Only TSE), further demonstrating the advantages of the joint model. The USEF-TP model achieves significantly higher SI-SDRi than other models across all overlap ratios, with a solid performance at high overlap rates (80\% and 100\%). It highlights the model’s robustness in scenarios with varying overlap ratios.

In summary, the USEF-TP model performs excellently in both SD and TSE tasks, exhibiting superior effectiveness and stability across different overlap conditions. Nonetheless, we observed a phenomenon where the model’s DER decreases as the overlap ratio of the test data increases. One possible reason is that, to compare with previous work, we used the same training and test sets as~\cite{maiti2023eend,Ao_2024}, namely LibriMix and SparseLibriMix. The LibriMix dataset mainly contains speech with high overlap ratios, likely leading to the model performing better on high-overlap speech. The data distribution of the SparseLibriMix test set might also be another factor contributing to this phenomenon. In future work, we plan to use LibriSpeech to simulate sparse overlapping speech closer to real-world data to reflect real-world scenarios better.
\begin{table}[]
\caption{Comparison of DER (\%) with previous models on the CALLHOME Dataset.}
  \label{tab:callhome}
\vspace{-0.3cm}
\centering
\begin{tabular}{c|ccccccc}
\hline \Xhline{1px}
\multirow{2}{*}{\textbf{Model}} & \multicolumn{6}{c}{\textbf{Number of Speakers}}                               \\
                                & \textbf{2} & \textbf{3} & \textbf{4} & \textbf{5} & \textbf{6} & \textbf{all} \\ \hline
AHC clustering~\cite{Horiguchi_2020}                  & 15.45      & 18.01      & 22.68      & 31.40      & 34.27      & 19.43        \\
Spectral clustering~\cite{wang2018speaker}             & 16.05      & 18.77      & 22.05      & 30.46      & 36.85      & 19.83        \\
EEND-EDA~\cite{Horiguchi_2020}                        & 8.50       & 13.24      & 21.46      & 33.16      & 40.29      & 15.29        \\
AED-EEND~\cite{chen2023attention}                        & 6.96       & 12.56      & 18.26      & 34.32      & 44.52      & 14.22        \\
TS-VAD~\cite{medennikov2020target}                          & 12.95      & 14.57      & 20.26      & 27.36      & 32.66      & 15.51        \\
\added{DiaPer~\cite{landini2024diaper}}                          & \added{7.39}       & \added{12.08}      & \added{19.62}      & \added{30.25}     & \added{28.84}      & \added{13.60}        \\
USED-F~\cite{Ao_2024}                          & 9.09       & 12.76      & 14.67      & 26.96      & 26.71      & 13.16        \\ 
\added{USED~\cite{Ao_2024}}                          & \added{10.23}       & \added{12.78}      & \added{14.68}      & \added{32.02}      & \added{25.10}      & \added{13.51}        \\\hline
USEF-TP                         & 9.21       & 14.43      & 20.48      & 32.96      & 26.27      & 15.26        \\ \hline \Xhline{1px}
\end{tabular}
\vspace{-0.2cm}
\end{table}
\subsection{Results on the CALLHOME Dataset}
\added{Table~\ref{tab:callhome} presents the DER comparison between the proposed USEF-TP model and several state-of-the-art SD systems on the CALLHOME dataset. As shown, USEF-TP achieves an overall DER of 15.26\%, which is competitive with models such as EEND-EDA (15.29\%) and TS-VAD (15.51\%) and outperforms traditional clustering-based methods like AHC and spectral clustering by a significant margin. This demonstrates the effectiveness of USEF-TP in real conversational scenarios.}

\added{However, the USEF-TP model still falls short compared to the recently proposed USED~\cite{Ao_2024} (13.51\%) and DiaPer~\cite{landini2024diaper} (13.60\%), particularly in scenarios with five speakers, where the performance gap becomes more evident. A potential reason is that conventional SD or speech separation models (e.g., DiaPer and USED) simultaneously generate outputs for all speakers in the mixed speech, inherently modeling the relationships among all speakers. In contrast, TSE or PVAD models (e.g., USEF-TP) focus only on a target speaker, producing predictions about the target speaker. The ability to capture speaker interactions in SD and speech separation models may be beneficial for improving performance in complex multi-speaker scenarios. In future work, we may leverage this aspect to enhance the model’s performance in real conversational scenarios.}

\section{Conclusion}
\label{sec:Conclusion}
This paper proposes a Universal Speaker Embedding Free Target speaker extraction and Personal voice activity detection (USEF-TP) model. The USEF-TP model performs target speaker extraction and personal voice activity detection jointly. USEF-TP utilizes an embedding-free framework as the network backbone, and we designed an interaction module to enable PVAD results to inform TSE predictions. Additionally, we introduced an energy loss to enhance the model’s performance across mixed speech with varying overlap ratios. Experimental results indicate that our proposed USEF-TP demonstrates superior performance and stability across overlapping conditions in PVAD and TSE tasks.

\section{Acknowledgments}
This research is funded in part by the National Natural Science Foundation of China (62171207), Science and Technology Program of Suzhou City(SYC2022051) and Guangdong Science and Technology Plan (2023A1111120012). Many thanks for the computational resource provided by the Advanced Computing East China Sub-Center.

\newpage
\bibliographystyle{elsarticle-num}
\bibliography{mybibfile}

\end{document}